\documentclass[oneside,11pt]{article}

\usepackage[utf8]{inputenc}
\usepackage[T1]{fontenc}
\usepackage{amsthm}
\usepackage{amsmath}
\usepackage{amsfonts}
\usepackage{amssymb}
\usepackage{eurosym}
\usepackage{bbm}
\usepackage{graphicx} 
\usepackage{float}
\usepackage{booktabs}
\usepackage{multirow}
\usepackage{rotating}
\usepackage{array}
\usepackage{xcolor}
\usepackage{subcaption}
\usepackage[ruled]{algorithm2e}
\newcommand{\lIfElse}[3]{\lIf{#1}{#2 \textbf{else}~#3}}
\newcolumntype{R}[1]{>{\raggedleft\let\newline\\\arraybackslash\hspace{0pt}}m{#1}}
\newcolumntype{L}[1]{>{\raggedright\let\newline\\\arraybackslash\hspace{0pt}}m{#1}}
\usepackage{breakcites}
\usepackage{nowidow}
\usepackage{enumitem}
\usepackage[top=1.5cm,bottom=2cm,right=2.5cm,left=2.5cm]{geometry}
\usepackage{titling}
\usepackage{natbib}
\usepackage{ifplatform}
\usepackage{textcomp}
\ifwindows
\fi
\allowdisplaybreaks

\newcommand{\R}{\mathbb{R}} % real number set
\newcommand{\N}{\mathbb{N}} % normal distribution
\newcommand{\cD}{\mathcal{D}} % stopping set
\newcommand{\cC}{\mathcal{C}} % continuation set
\newcommand{\sgn}{\mathrm{sgn}} % continuation set
\newcommand{\Ind}{\mathbbm{1}} % indicator function
\newcommand{\lp}{\left(} % left parenthesis
\newcommand{\rp}{\right)} % right parenthesis
\newcommand{\lc}{\left\{} % left curly brackets
\newcommand{\rc}{\right\}} % right curly brackets

\newcommand{\wh}[1]{\widehat{#1}}
\newcommand{\ol}[1]{\overline{#1}}
\newcommand{\ul}[1]{\underline{#1}}
\newcommand{\lrp}[1]{\lp#1\rp} % left-right parenthesis
\newcommand{\lrc}[1]{\lc#1\rc} % left-right curly brackets
\newcommand{\Prob}[1]{\mathbb{P}\lp #1\rp} % probability 1
\usepackage[pdftex,final,bookmarksnumbered,bookmarksopen=false,breaklinks,colorlinks]{hyperref}
\hypersetup{
	final,
	bookmarksnumbered=true,
	bookmarksopen=true,
	bookmarksopenlevel=0,
	unicode=false,
	pdftoolbar=true,
	pdfmenubar=true,
	pdffitwindow=false,
	pdftitle={Optimal stopping of Gauss-Markov bridges},
	pdfdisplaydoctitle=true,
	pdftoolbar=true,
	pdfmenubar=true,
	pdflang={English},
	pdfauthor={Abel Azze, Guglielmo D'Amico, Bernardo D'Auria, Salvatore Vergine},
	pdfsubject={arXiv paper},
	pdfcreator={Abel Azze, Salvatore Vergine},
	pdfproducer={Abel Azze, Salvatore Vergine},
	pdfkeywords={Brownian bridge}{Monte Carlo simulation}{Power ramping}{Semi-Markov process }{Wind energy},
	pdfnewwindow=true,
	breaklinks=true,
	hidelinks,       
	linkcolor=black,
	citecolor=black,
	filecolor=magenta,
	urlcolor=cyan
}

\newif\ifmain
\maintrue
\newif\ifsupplement
\supplementtrue

\newif\iffigstabs
\figstabstrue

\begin{document}

\ifmain

%-----------------------------------------------%
\title{Modelling a storage system of a wind farm with a ramp-rate limitation: a semi-Markov modulated Brownian bridge approach}
\setlength{\droptitle}{-1cm}
\predate{}%
\postdate{}%
\date{}
%-----------------------------------------------%

%-----------------------------------------------%
\author{Abel Azze$^{1}$, Guglielmo D'Amico$^{2}$, Bernardo D'Auria$^{3}$, and Salvatore Vergine$^{4, 5}$}
\footnotetext[1]{Department of Quantitative Methods, CUNEF Universidad (Spain).}
\footnotetext[2]{Department of Economics, University G. d’Annunzio of Chieti--Pescara (Italy).}
\footnotetext[3]{Department of Mathematics ``Tullio Levi Civita'', University of Padova (Italy).}
\footnotetext[4]{Department of Management, Marche Polytechnic University (Italy)}
\footnotetext[5]{Corresponding author. e-mail: \href{mailto:s.vergine@staff.univpm.it}{s.vergine@staff.univpm.it}.}
\maketitle
%-----------------------------------------------%

\begin{abstract}
    We propose a new methodology to simulate the discounted penalty applied to a wind-farm operator by violating ramp-rate limitation policies. 
    It is assumed that the operator manages a wind turbine plugged into a battery, which either provides or stores energy on demand to avoid ramp-up and ramp-down events. 
    The battery stages, namely charging, discharging, or neutral, are modeled as a semi-Markov process. 
    During each charging/discharging period, the energy stored/supplied is assumed to follow a modified Brownian bridge that depends on three parameters. 
    We prove the validity of our methodology by testing the model on 10 years of real wind-power data and comparing real versus simulated results.
\end{abstract}
\begin{flushleft}
\end{flushleft}

%-------------------------------------------------%
\section{Introduction}\label{sec:intro}
%-------------------------------------------------%

% Introduction
In the last decades, we have assisted in the increase of renewable energy penetration in the electricity market, in particular from wind and solar sources. This is caused by the increasing concern about environmental pollution and global warming, and the awareness of having to exploit sources of clean energy to decrease the use of fossil fuels \citep{razmjoo2021technical, janzen2020greenhouse,biancardi2023flexibility}.

% The problem of intermittent energy production
One of the main problems that hinder the use of wind power is its intermittent nature caused by rapid and unpredictable fluctuations in wind speed. This conflicts with the stability required by the energy market to guarantee a systemic balance and security \citep{frate2020impact}.

% Introduction to ramp-rate limitations
Among the control strategies used to decrease the high variability of wind power, the ramp-rate limitation has seen increasing use in recent years \citep{hittinger2014effect,bossavy2015edge,d2021analysis,d2022ramp}. Limiting the ramp rate means limiting the rate at which the power production varies between two consecutive time steps.
The ramp-rate limits might be violated in two ways: up-ramping events, meaning that the variation is positive, and down-ramping events, when the change is negative \citep{gallego2015review}. Predicting these two types of events have been in the spotlight of wind-farm managing literature for a while \citep{cui2021algorithm,zheng2022offshore}.

The ramp-rate event is considered a critical event because its delayed and inadequate control can cause serious damage to the power grid and consequent economic losses. Several of the largest system operators, such as the Electric Reliability Council of Texas (ERCOT) and the state-owned electric power transmission operator in Ireland, EirGrid, require ramp-rate control to wind generators. According to the grid characteristics, it can be requested to control the ramp rate within one- and ten-minutes limits \citep{cui2023algorithm,hittinger2014effect,d2022ramp}.

% Batteries as a solution to avoid ramp-rate violations
Wind farms subjected to ramp-rate limitations usually use a storage system for two main purposes: providing power when a ramp-down occurs, and storing power in the presence of a ramp-up event \citep{hittinger2010compensating,teleke2009control,lone2008modelling,khalid2010model,abdullah2014effective}. Batteries are the most used energy storage systems due to their quick response time and easy installation, and, in this context, its main variables of interest are the size and the state of charge, but in principle also pumped-storage generating system can be used \citep{li2019dynamic}. 

In literature, particular attention is given to the ramp rate detection and prediction (\cite{cui2021algorithm,zheng2022offshore}), and, more in general, to the application of stochastic processes that make the operator able to know in advance the future behaviour of the system in terms of wind power variability \citep{lee2012limiting,d2013first,chen2009arima,an2012short,grassi2010wind}. These aspects are relevant from an economic perspective due to the possibility of forecasting the economic losses of a wind farm under a ramp-rate policy
\citep{cui2021algorithm,cui2017data}. The amount of penalty can be consistent if we consider that, for example, that in ERCOT the penalty is computed by multiplying the energy (in MW-h) not meeting the ramp-rate up or down limitations by the regulation up or down prices, which were $16\$$/MW-h and $13\$$/MW-h in 2008–2009, respectively \citep{hittinger2014effect}. According to \cite{wan2011analysis}, in Texas, between 2004 and 2009, the number of large ramp events with magnitude $>25\%$ of the highest annual wind generation is $235$ per year, on average. If we want to have a rough estimate, we can assume that the average installed capacity in Texas is $2500$ MW in such a period, and the average amount of ramp events with magnitude $>25\%$ of the highest annual wind generation is 625 MW. By considering an average penalty of $14.5\$$/MW-h, we obtain a total average penalty of about $9000,00\$$ per year. %Clearly, 
These economic losses refer to the percentage of $>25\%$, which is very high. If we lower the percentage the losses increase.

% Introducing the model of the battery operations, charges, and the penalty process
Ramp-rate limitations are usually coupled with a penalty policy if the wind farm does not meet the imposed limits \citep{hittinger2014effect}. Among the existing penalty systems applied by the system operators, one largely used method consists of multiplying a fixed monetary amount and the number of MW above or below the ramp-rate limits. The work of \cite{d2021analysis} considers a system comprising a wind farm with a ramp-rate limitation policy and a battery, with the aim of forecasting the penalties received by the operator over a given time period. This kind of system shows a nonlinear behaviour, which is due to the interaction between the charge and discharge processes and the storage capacity of the battery. Indeed, as ramp events occur throughout time, the battery's state of charge shifts accordingly. This affects the ability to dispatch or store energy as needed to avoid penalties. In summary, the variations of energy are subject to a time-varying nonlinear random constraint, which is the result of the wind speed fluctuations, state of charge, and ramp-rate policy. A discrete-time homogeneous Markov chain is used to model the battery operations, which are divided into three states: the charging event, the discharging event, and the neutral event or absence of operations.
During each charging/discharging period, the random power stored/supplied by the battery is assumed to be a discrete collection of independent, not identically distributed, random variables.
The penalty is then calculated by multiplying the random charges/discharges by the regulation fees.
However, the Markovian assumption was proved to be not completely satisfactory in this context by \cite{d2022modelling}, where a semi-Markov process was instead considered to model the battery operations and, consequently, the penalty dynamics were set to evolve as a semi-Markov modulated reward process. This kind of stochastic process has been largely used in the literature (\cite{d2013wind,feinberg1994constrained,papadopoulou2012moments}).

% Contributions
This paper builds on the semi-Markov approach used in \cite{d2022modelling} by using a modification of a Brownian Bridge (BB) to model the charging/discharging processes during the battery's operation periods. Hence, we take off the independence assumption, considered by \cite{d2021analysis} and \cite{d2022modelling}, between the random charges/discharges at different times, which does not seem to be supported by real data and can be considered only as an approximation of the behaviour of the system. Besides, a BB model accounts for the convenient Markovian and Gaussian properties along the waiting time of the underlying semi-Markov process, and it is one of the most exhaustively studied diffusion bridges, making it an appealing model from theoretical and applicable perspectives.
We use 10 years of wind speed real data to compute the power production of a hypothetical wind turbine located in Sardinia, which we use to obtain the penalty associated to three different ramp-rate limitations: $1\%$, $5\%$ and $7\%$ of the wind turbine rated capacity. An estimation of the penalty process is then produced via a Monte Carlo simulation algorithm. 
The results suggest that our semi-Markov-modulated model succeeds in simulating the accumulated penalty process over a given time period.

% Structure of the paper
The rest of the paper is structured as follows. Section \ref{sec:RR_sM} introduces the ramp-rate policy and sets the semi-Markov model that governs the battery operation process. When the battery is either charging or discharging, a discrete-time model based on a BB is proposed in Section \ref{sec:GM_model} to model the dynamics of the random charges.
We then introduce, in Section \ref{sec:SoC_penalty}, two key processes: the one associated with the state of the charge of the battery, and the penalty process. The last one is then generated via Monte Carlo simulations in Section \ref{sec:simulations}, where we show the competitiveness of our method by comparing its results against the penalty obtained from real data. Final thoughts are relegated to Section \ref{sec:conclusions}.

%-------------------------------------------------%
\section{The semi-Markov model of sequential ramp-rate events}\label{sec:RR_sM}
%-------------------------------------------------%

Ramp-rate limitations can be used to smooth the power produced from the wind turbine and obtain a more stable output. We limit both up-ramping and down-ramping events, as done in \cite{hittinger2014effect}. This limitation decreases the slope by which the power profile changes between two consecutive time steps, it is indicated as a percentage of the rated capacity of the wind farm, and its unit of measure can be MW/h. Lower percentages represent stricter limitations. For example, if we consider a rated capacity of $2$ MW, a ramp-rate limitation of $1\%$ penalizes changes faster than $0.02$ MW/h.

For $\delta > 0$ and $k\in\N$ let $e(k)$ be the power generated at time $\delta k$, and define the modified power at the same time, denoted by $\bar{e}(k)$, as
\begin{equation}
\bar{e}(k) := \left\{
\begin{array}{rl}\label{eq:corrected_power}
\bar{e}(k-1) + \ell\delta, & \text{if } e(k) > \bar{e}(k-1) + \ell\delta \quad\quad\quad \text{(up-ramping event)} \\
\bar{e}(k-1) - \ell\delta, & \text{if } e(k) < \bar{e}(k-1) - \ell\delta \quad\quad\quad \text{(down-ramping event)} \\
e(k), & \text{otherwise }
\end{array}
\right.
\end{equation}
where $\ell > 0$ is the ramp-rate limit. For the sake of simplicity and because it does not sacrifice generality, we set $\delta = 1$ for the rest of the paper. Working with an arbitrary time-step length follows identical steps. The following is an interpretation of the earlier quantities: the power generated by the wind turbine at time $k$ is denoted by $e(k)$, while the power that is effectively injected into the electrical grid at time $k$ is denoted by $\overline{e}(k)$. The ramp-rate policy is employed to stabilize the grid because $\overline{e}(k)$ exhibits milder variations and less steep slopes than $e(k)$.

We connect a battery to the wind turbine to either store or supply energy in case of an up-ramping or down-ramping event, respectively. A penalization is applied every time the battery cannot store the total energy surplus or provide the required energy.
 
We consider the battery operations over time as a Markov chain $\lrc{J_n}_{n\in \N}$ with state space $E=\{-1,0,+1\}$. States $-1$ and $+1$ indicate discharging and charging operations, associated, respectively, with a down-ramping and up-ramping event.
The state $0$ represents the unchanged condition, that is, the battery is neither charging nor discharging, which occurs when the power production meets the ramp-rate limits. 
The process $\lrc{K_n}_{n\in \N}$ stands for the $n$-th time in which the battery changes state. We assume that $K_0 = 0$ and $K_n < K_{n+1}$ for all $n \in\N$. The (sojourn) time the battery remains in the state $J_{n}$, before the $(n+1)$th jump, is denoted by $X_{n+1}$. Formal definitions of $J_n$, $K_n$, and $X_n$, are given below:
\begin{align*}
    K_n &= \inf\lrc{k > K_{n-1} : \sgn(e(k) - \bar{e}(k)) \neq \sgn(e(K_{n-1}) - \bar{e}(K_{n-1}))}, \\
    J_n &= \sgn(e(K_n) - \bar{e}(K_n)), \\
    X_{n+1} &= K_{n+1} - K_{n},\quad  X_{0}=0.
\end{align*}

We assume that $\lrc{(J_n, K_n)}_{n\in \N}$ is a Markov Renewal process, and define its kernel as
\begin{align*}
q_{i,j}(k) &:= \Prob{{J}_{n+1} = j, {X}_{n+1} = k \mid {J}_{n} = i} \\
&= \Prob{{J}_{n+1}=j,{X}_{n+1} = k \mid {J}_{n} = i, J_{n-1}, \dots, {J}_{0}, {K}_{n}, \dots, {K}_{0}}.
\end{align*}
According to the previous relation, regardless of what the values of the past variables were, knowing the last battery operation, $J_n$, is sufficient to provide the conditional joint distribution of the pair, $J_{n+1}$, and $X_{n+1}$.

For later reference, we introduce the conditional sojourn-time distribution
\begin{align}\label{eq:sojourn-time_density}
    h_i(k) := \Prob{X_{n+1} = k\ |\ J_n = i}
    = \sum_{j \in E} q_{ij}(k)    
\end{align}
as well as the transition probabilities of the embedded Markov chain
\begin{align}\label{eq:transition_prob}
    p_{ij} := \Prob{J_{n+1} = j\ |\ J_n = i}
    = \sum_{k = 0}^{\infty} q_{ij}(k),   
\end{align}
and the conditional (to the sojourn time) transition probabilities
\begin{align}\label{eq:conditional_transition_prob}
    p_{ij}(k) &:= \Prob{J_{n+1} = j | J_n = i, X_{n+1} = k} = q_{ij}(k)/h_i(k).   
\end{align}

Finally, consider the process $N(k) = \max\{{n > 0:K_n \leq k\}}$, which counts the number of transitions up to time $k$, as well as the Semi-Markov Chain (SMC) associated with the Markov Renewal chain $(J_{n}, K_{n})$, indicated by $Z(k) := J_{N(k)}$, for $k\in\N$.
%-------------------------------------------------%
\section{A Gauss--Markov model for battery operations between ramp-rate events}\label{sec:GM_model}
%-------------------------------------------------%

The purpose of this section is to model the battery storing operations using a BB, which is a well-studied process and accounts for two desirable properties, namely Gaussianity and Markovianity. 

\subsection{Setting the model}\label{subsec:model}

Consider the semi-Markov model of battery operations introduced in the previous section and call a random segment to any triplet $(J_n, J_{n+1}, X_{n+1})$, for $n\in\N$. Hence, a random segment comprises an initial state $J_n$ denoting the current operation, its time length $X_{n+1}$, and the next operation $J_{n+1}$. Thereby, the triplet $(J_n = +1, J_{n+1} = -1, X_{n+1} = 5)$, for instance, denotes a segment where the battery after charging for $5$ units of time starts discharging afterwards.

Let $C(k)$ represents the theoretical random energy charged/discharged into/from the battery whenever possible at time $k\in\N$. Define $C_{J_n, J_{n+1}, X_{n+1}} := \lrc{C_{J_n, J_{n+1}, X_{n+1}}(k)}_{k = 1}^{X_{n+1}}$ be the process representing the charge ($J_n = +1$) or discharge ($J_n = -1$) during the $n$th stage of the SMC. Note that this notation presumes that the charging/discharging process depends on $J_n$, $J_{n+1}$, and $X_{n+1}$, and is independent of $K_n$. We assume that it is also independent of $\lrc{C(m)}_{m = 0}^{K_n - 1}$. That is, for $k = 1, \dots, x$,
\begin{align*}
    \Prob{C(K_n + k - 1) \leq c\ \big|\ J_n = i, J_{n+1} = j, X_{n+1} = x, K_n = K, \lrc{C(m)}_{m = 0}^{K_n - 1}} = \Prob{C_{i, j, x}(k) \leq c}.
\end{align*}

Consider the discrete-time processes $\cC_{i,j,x} := \lrc{\cC_{i, j, x}(k)}_{k = 0}^{x+1}$, for $i\in \{-1, +1\}$, $j \in \{-1, 0, +1\}\backslash\{i\}$, $x \in \N$.

Moreover, set
\begin{enumerate}[leftmargin=1.2cm,label=(P.\textit{\arabic{*}}),ref=P.\textit{\arabic{*}}]
    \item $\lrc{\cC_{i,j,x}(k)}_{k = 1}^x = \lrc{|C_{i,j,x}(k)|}_{k = 1}^x$, \label{P:C_n=C}
    \item $\cC_{i,j,x}(0) = \cC_{i,j,x}(x+1) = 0$. \label{P:C_n=0}
\end{enumerate}
In this way, we are embedding the charging process $C_{i,j,x}$ into the bridge process $\cC_{i,j,x}$. Indeed, \eqref{P:C_n=C} sets the embedding while \eqref{P:C_n=0} ensures that $\cC_{i,j,x}$ can be regarded as a bridge that vanishes at its initial and terminal times. Note that the absolute-value transformation in condition \eqref{P:C_n=C} does not introduce identifiability issues as $C_{i,j,x}$ has constant sign. Figure \ref{fig:power-cpower} shows some linearly-interpolated sample paths of $\cC_{i,j,x}$.

\begin{figure}[!ht]
	\centering
        \begin{subfigure}[b]{0.49\textwidth}
		\includegraphics[width = \textwidth]{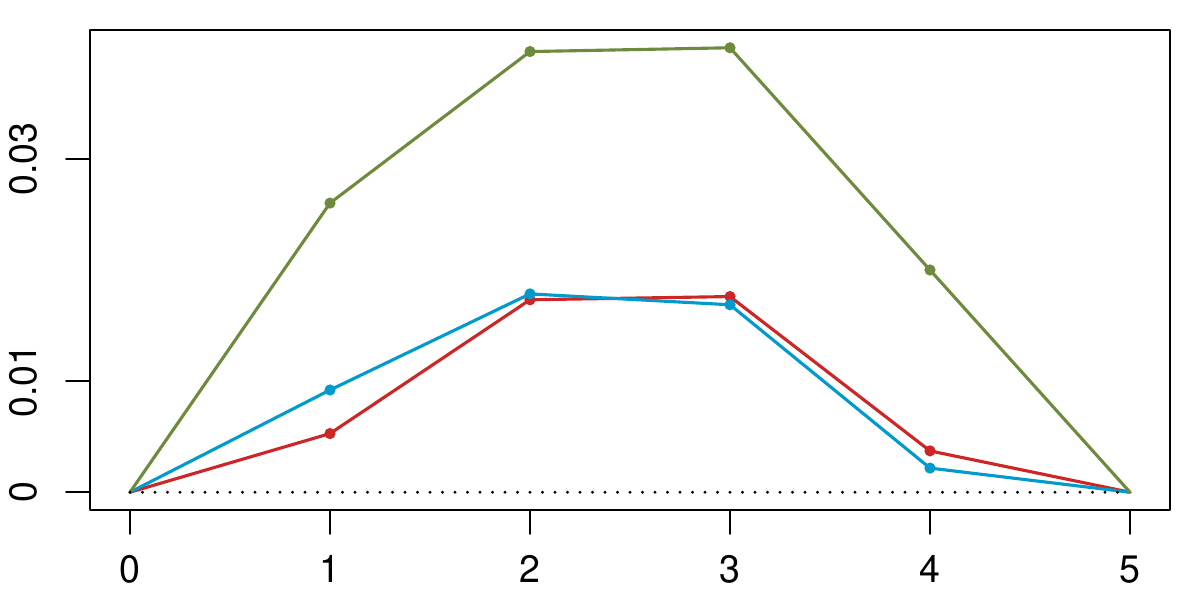}
		\subcaption{}
	\end{subfigure}
	\begin{subfigure}[b]{0.49\textwidth}
		\includegraphics[width = \textwidth]{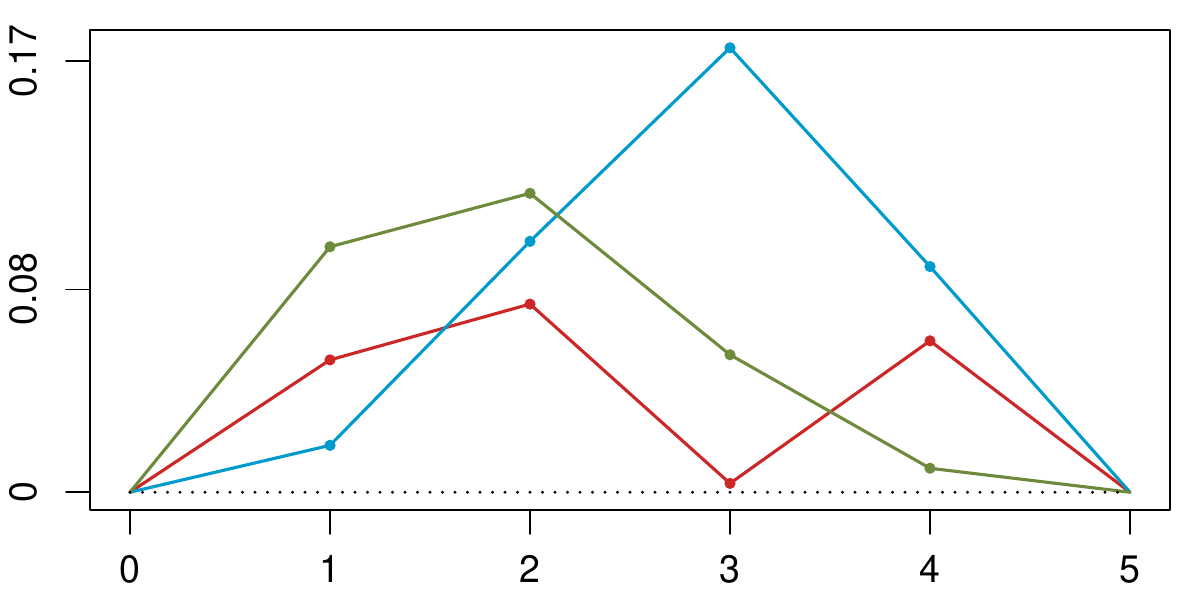}
		\subcaption{}
	\end{subfigure}
	\caption{Battery charges/discharges $\cC_{i, j, x}$. The colored dots represent the actual values of the charges/discharges, which are linearly interpolated. Figure (a) accounts for $(i = -1, j = 0, x = 4)$, while $(i = +1, j = 0, x = 4)$ in figure (b).}
	\label{fig:power-cpower}
\end{figure}

Essentially, properties \eqref{P:C_n=C} and \eqref{P:C_n=0} define the size of the  charging/discharging values for each segment of the SMC that is enlarged by introducing two boundary conditions at the times $0$ and $T_n$, where the size of the charging/discharging is set to zero.

For a neater notation, we will define the shorthands $C_n := C_{J_n, J_{n+1}, X_{n+1}}$ and $\cC_n := \cC_{J_n, J_{n+1}, X_{n+1}}$.

We introduce now a parsimonious model for $\cC_{i,j,x}$. Define the parameters
\begin{align}\label{eq:max_parameter}
    \tau_{i,j,x} &:= \arg\max\lrc{\cC_{i,j,x}(k) : k = 1,\dots, x},\quad h_{i,j,x} := \cC_{i,j,x}(\tau_n),
\end{align}
and the error process
\begin{align}\label{eq:error-process}
    E_{i,j,x}(k) := \cC_{i,j,x}(k) - g_{i,j,x}(k),
\end{align}
where, for $0 < \tau < x+1$ and $h > 0$, $g_{i, j , x}$ is given by 
\begin{align}\label{eq:triangle}
    g_{i, j, x}(t) &:= \frac{h_{i,j,x}}{\tau_{i,j,x}}t\Ind(t\leq \tau_{i,j,x}) + \frac{h}{x+1 - \tau_{i,j,x}}(x+1 - t)\Ind(t > \tau_{i,j,x})
\end{align}
Thus, $C_n$ can be split into the sum of the (triangle-shape) function $g_{i,j,x}$ plus the stochastic process $E_{i,j,x}$. Figure \ref{fig:charge_decomposition} shows a few paths of these three processes.

\begin{figure}[!ht]
	\centering
	\begin{subfigure}[b]{0.32\textwidth}
		\includegraphics[width = \textwidth]{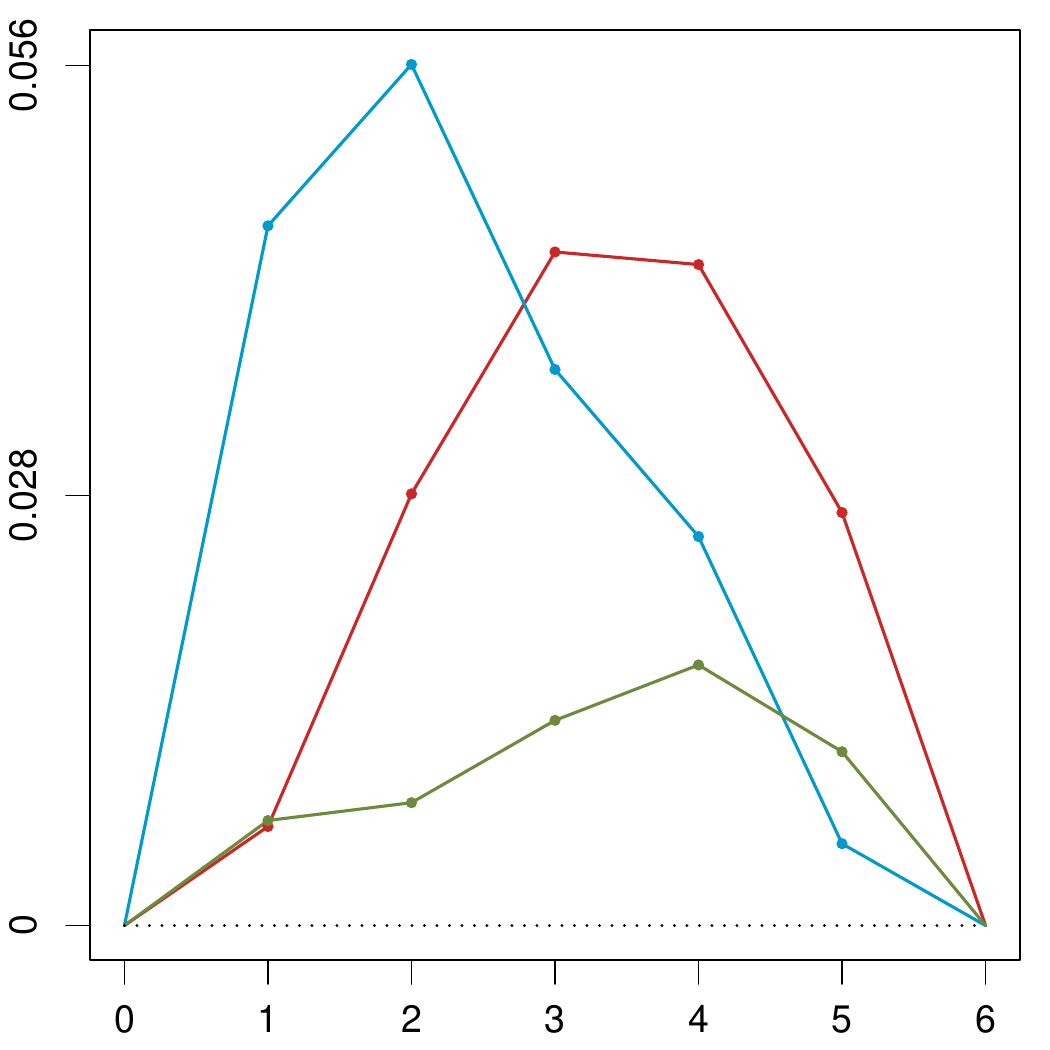}
		\subcaption{$\cC_{i, j, x}$}
	\end{subfigure}
	\begin{subfigure}[b]{0.32\textwidth}
		\includegraphics[width = \textwidth]{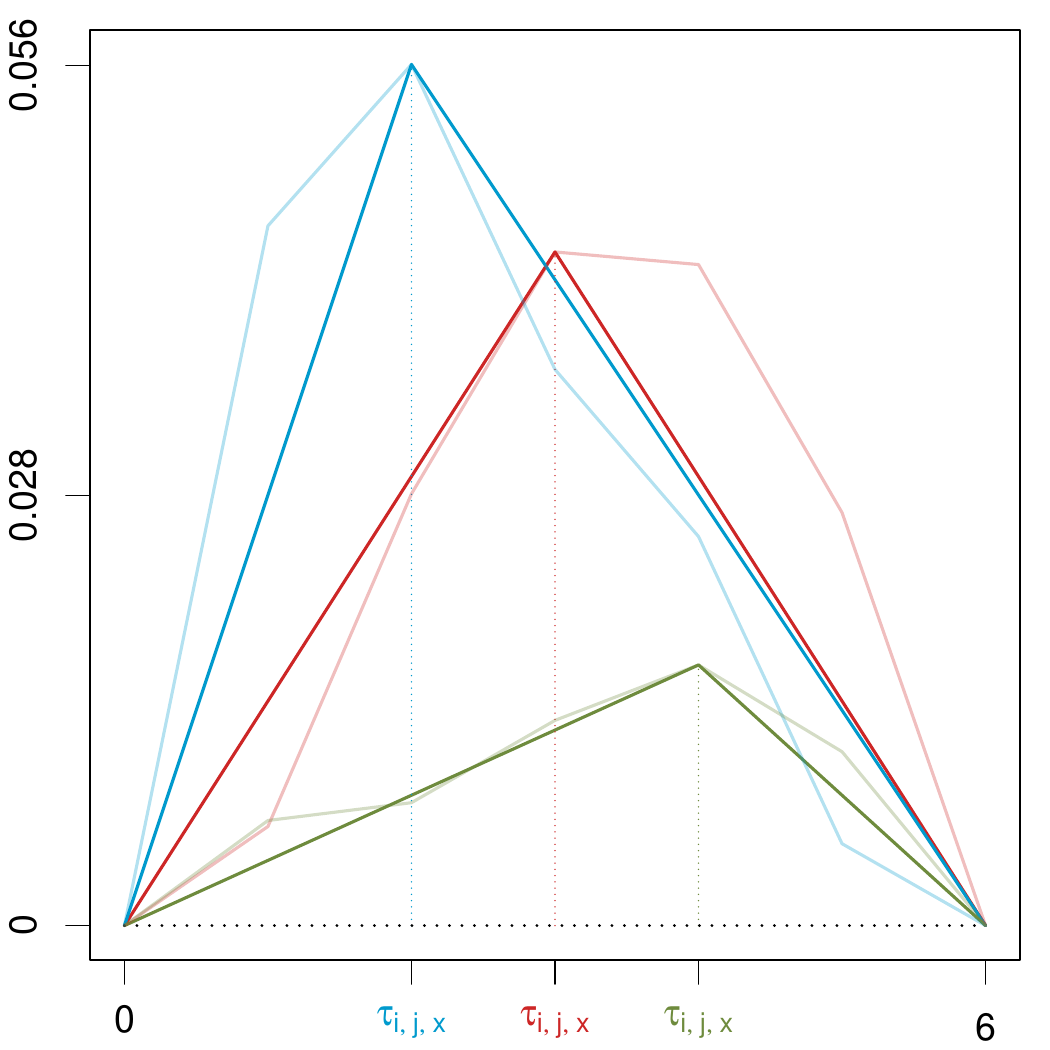}
		\subcaption{$g_{i, j, x}$}
	\end{subfigure}
	\begin{subfigure}[b]{0.32\textwidth}
		\includegraphics[width = \textwidth]{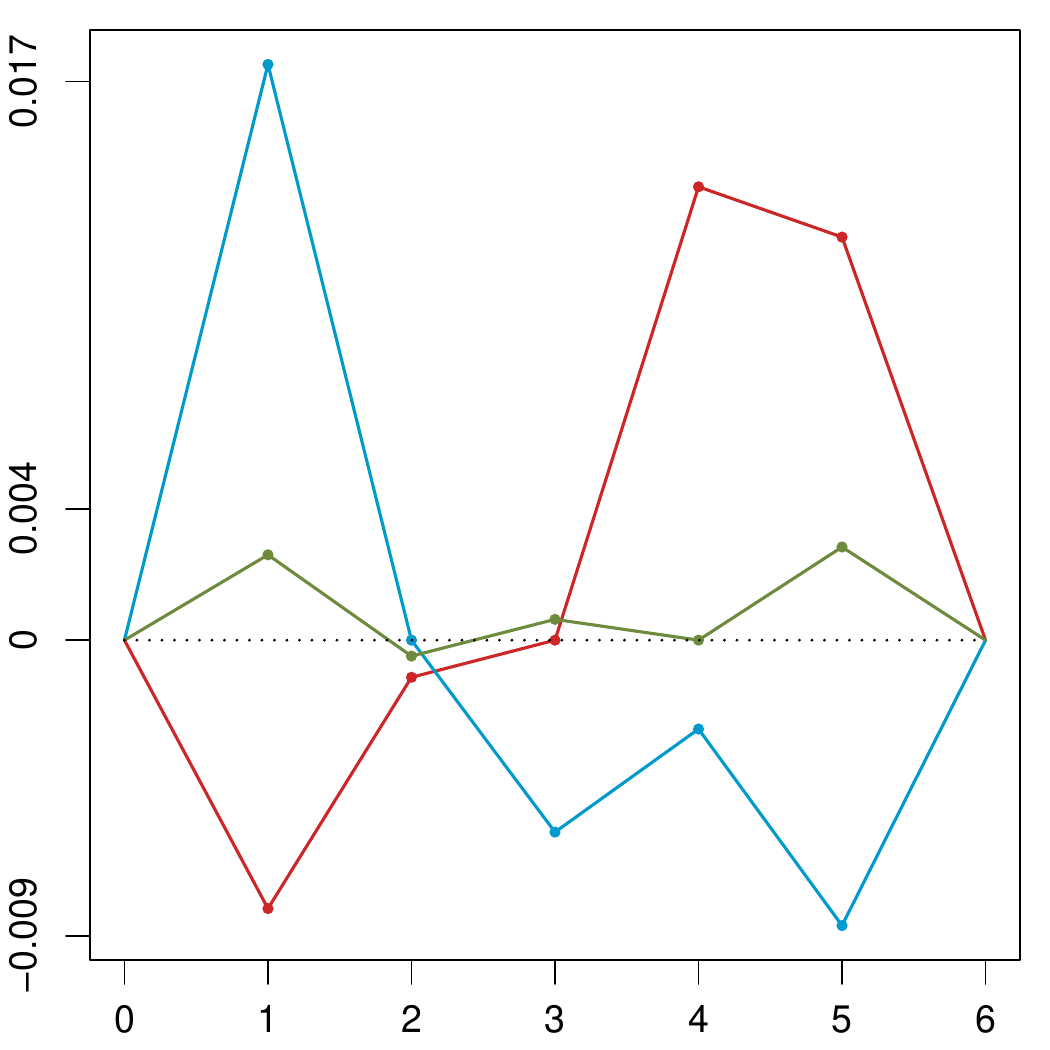}
		\subcaption{$E_{i, j, x}$}
	\end{subfigure}
	\vspace{0.2cm}
	\caption{Image (a) shows three paths of the bridge process $\cC_{i, j, x}$. Image (b) displays these same charges (transparent lines), alongside the triangle functions $g_{i, j, x}$ derived from them. The associated error process $E_{i, j, x}$ is shown in image (c).  All images account for $i = -1$, $j = 0$, $x = 5$, and the ramp-rate coefficient $\ell = 0.02$. The values of $\{\cC_{i, j, x}(k)\}_{k = 1}^x$ are remarked using bullet points in the curves that result after linearly interpolating them.}
    \label{fig:charge_decomposition}
\end{figure}

Let $\bar{e}_{i, j, x}$ be the corrected power at the time when the battery changes to the state $i\in E$ and remains there for a sojourn time $x$, after which it changes to the state $j \in E\backslash\{i\}$. Hence, $\bar{e}_{i, j, x}$ has the same distribution as $\bar{e}(K_n)$ for all $n$ such that $(J_n = i, J_{n+1} = j, X_{n+1} = x)$. We are implicitly assuming that $\bar{e}(K_n)$ depends entirely on $J_n$, $J_{n+1}$, and $X_{n+1}$.

Consider now the variable
\begin{align}\label{eq:initial_power}
    \rho_{i, j, x} := 
    \begin{cases}
        \min\{\max\{\bar{e}_{i,j,x} - \ell, \ell (x+1)\}, P_r\}, & \text{ if } i = -1 \\
        \max\{P_r - (\bar{e}_{i,j,x} - \ell), P_r - \ell (x+1), 0\},  & \text{ if } i = +1
    \end{cases},
\end{align}
where $P_r$ is the rated capacity of the wind turbine.
The variable $\rho_{i, j, x}$ can be regarded as the initial corrected power along the random segment $(J_n = i, J_{n+1} = j, X_{n+1} = x)$. 

In \eqref{eq:initial_power}, the adjustments $\bar{e}_{i,j,x} - \ell$ and $Pr -(\bar{e}_{i,j,x}) - \ell$ are justified by the assumption that the battery has been charging/discharging for exactly one unit of time before the first observed battery operation of the random segment. The lower bounds $\ell (x+1)$ and $Pr - \ell (x+1)$ are needed because the battery remains in the state $i$ exactly one unit of time after the last observed operation, making a total charging/discharging time length of $x+1$. Finally, the formula takes into account the fact that the initial power cannot exceed the rated capacity $P_r$ neither be negative.

Note that
\begin{align}\label{eq:charge-upper-limit}
    0 \leq \cC_{n}(k) = |(e - \bar{e})(K_n + k - 1)| \leq \rho_{J_n, J_{n+1}, X_n} - (k-1)\ell, \quad k = 1, \dots, x.
\end{align}
Inequality \eqref{eq:charge-upper-limit} alongside the non-negativity of $\cC_n$ yield the following restrictions to the error process $E_{i, j , x}$:
\begin{align}\label{eq:error-process-representation}
    E_{i, j, x}(k) = \max\lrc{-g_{i, j, x}(k), \min\lrc{Y_{i, j, x}(k), \rho_{i, j, x} - (k - 1)\ell - g_{i, j, x}(k)}}, \quad k = 1, \dots, x,
\end{align}
for some process $Y_{i, j, x} := \lrc{Y_{i, j, x}(k)}_{k = 1}^x$. We choose to model $Y_{i, j, x}$ as a BB going from $(0, 0)\in \mathbb{R}^2$ to $(T, 0)\in \mathbb{R}^2$ and forced to stop by $(\tau_{i, j, x}, 0)$. Hence, $Y_{i, j, x}$ admits the representation 
\begin{align}\label{eq:BB}
    Y_{i, j, x}(k) = Y_{i, j, x}^{(1)}(k \wedge \tau_{i, j, x}) + Y_{i, j, x}^{(2)}((k - \tau_{i, j, x})\vee 0), \quad k = 1, \dots, x,
\end{align}
where $Y_{i, j, x}^{(1)}$ and $Y_{i, j, x}^{(2)}$ are two independent BBs satisfying the representation
\begin{align}\label{eq:BB1}
    Y_{i, j, x}^{(1)}(k) &:= \sigma_{i, j , x}\lrp{W^{(1)}(k) - \frac{k}{\tau_{i, j, x}}W^{(1)}(\tau_{i, j, x})}, \quad k = 1, \dots, x, 
\end{align}
and
\begin{align}\label{eq:BB2}
    Y_{i, j, x}^{(2)}(k) &:= \sigma_{i, j, x}\lrp{W^{(2)}(k) - \frac{k}{x+1 - \tau_{i, j, x}}W^{(2)}(x+1 - \tau_{i, j, x})}, \quad k = 1, \dots, x,
\end{align}
where $W^{(1)}$ and $W^{(2)}$ are independent standard Brownian motions, and $\sigma_{i, j , x} > 0$ is the common volatility term.

\subsection{Estimation of the parameters}\label{subsec:parameters}

We provide here a mechanism to generate the parameters $\rho_{i, j, x}$, $\tau_{i, j, x}$, $h_{i, j, x}$, and $\sigma_{i, j, x}$. Define the shorthand notations $\rho_n := \rho_{J_n, J_{n+1}, X_{n+1}}$, $\tau_n := \tau_{J_n, J_{n+1}, X_{n+1}}$, and $h_n := h_{J_n, J_{n+1}, X_{n+1}}$. 
For $\rho_n$, $\tau_n$, and $h_n$, in alignment to \cite{d2022modelling}, we assume that all their values belong to the same population as soon as they share the same sojourn time as well as initial and next charging stages. That is, if we consider a generic segment identified by the triplet $(J_n = i, J_{n+1} = j, X_{n+1} = x)$, then the set of values
\begin{align*}
    \lrc{(\rho_{m}, \tau_m, h_m) : J_m = i, J_{m+1} = j, X_m = x}_{m\in\N} 
\end{align*}
are simulations of the joint distribution of $(\rho_{i, j , x}, \tau_{i, j , x}, h_{i, j , x})$, which we denote by $f_{i, j , x}:\cD_n\mapsto\R_+$, and has the support
\begin{align*}
    \cD_{i, j , x} := \lrc{(\rho, \tau, h) : \rho \in [\ul{\rho}, \ol{\rho}],\; \tau \in \lrc{1, \dots, x},\; h \in (0, \ol{h})},
\end{align*}
with, in alignment to \eqref{eq:initial_power},
\begin{align*}
    \ul{\rho} =
    \begin{cases}
         0, & \text{if } i = +1 \\
         P_r - (x+1)\ell, & \text{if } i = -1
    \end{cases}, \quad
    \ol{\rho} =
    \begin{cases}
         (x+1)\ell, & \text{if } i = +1 \\
         P_r, & \text{if } i = -1
    \end{cases},
\end{align*}
and
\begin{align*}
    \ol{h} =
    \begin{cases}
        \rho - \tau\ell, & \text{if } i = +1 \\
        P_r - (\rho + \tau\ell ), & \text{if } i = -1
    \end{cases}.   
\end{align*}
The upper bound $\ol{h}$ comes after \eqref{eq:initial_power} and the definition of $h_n$ in \eqref{eq:max_parameter}.

We estimate $f_{i, j, x}$ in a non-parametric fashion by relying on kernel estimations of their vine copulas. This method is well-documented in \cite{Nagler_2014}. Besides the flexibility of its non-parametric nature, the main drive for choosing this technique was its robustness in high-dimensional frameworks.

This method simulates $\tau_{i, j, x}$ as a continuous variable within the interval $[0, x]$, and then we replace that original simulation with the nearest value within the support $\lrc{1, \dots, x}$. 

We leaned on the approach suggested by \cite{Geenens_2017} for the copulas density estimation. They build on a larger body of works that transform observations in the unit square $[0, 1]^2$ into $\R^2$, where standard kernel density estimation techniques can be used, and a back-transformation recovers the copula density estimation. Specifically, \cite{Geenens_2017} propose a local likelihood estimator by means of quadratic polynomials approximations. 

We performed the kernel density estimation with these specifications via the \texttt{R} package \texttt{kdevine} \citep{kdevine_package}. 

To avoid running into small data issues for the bandwidth matrix estimation, we bootstrapped the sample of any random segment with less than $10$ observations and introduced small perturbations to guarantee differences among the new data values.

We illustrate in Figure \ref{fig:parameters} how this method of estimating $\rho_{i, j, x}$, $\tau_{i, j, x}$, and $h_{i, j, x}$ captures the distribution of the real data.

\begin{figure}[!ht]
	\centering
	\begin{subfigure}[b]{0.49\textwidth}
		\includegraphics[width = \textwidth]{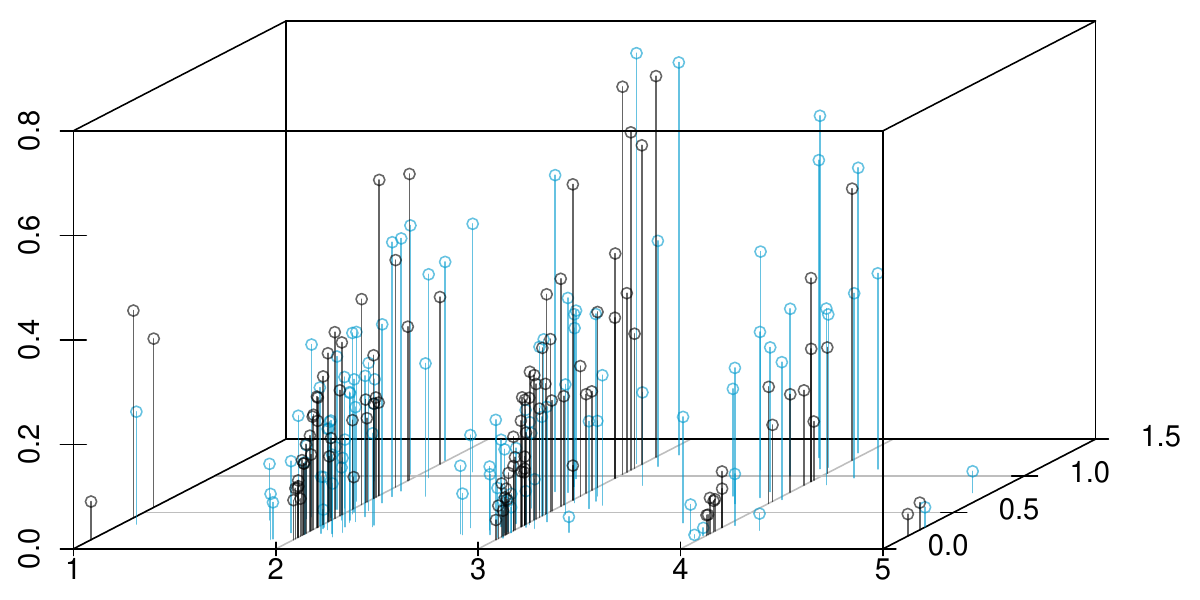}
		\subcaption{$X_{n+1} = 5$}
	\end{subfigure}
	\begin{subfigure}[b]{0.49\textwidth}
		\includegraphics[width = \textwidth]{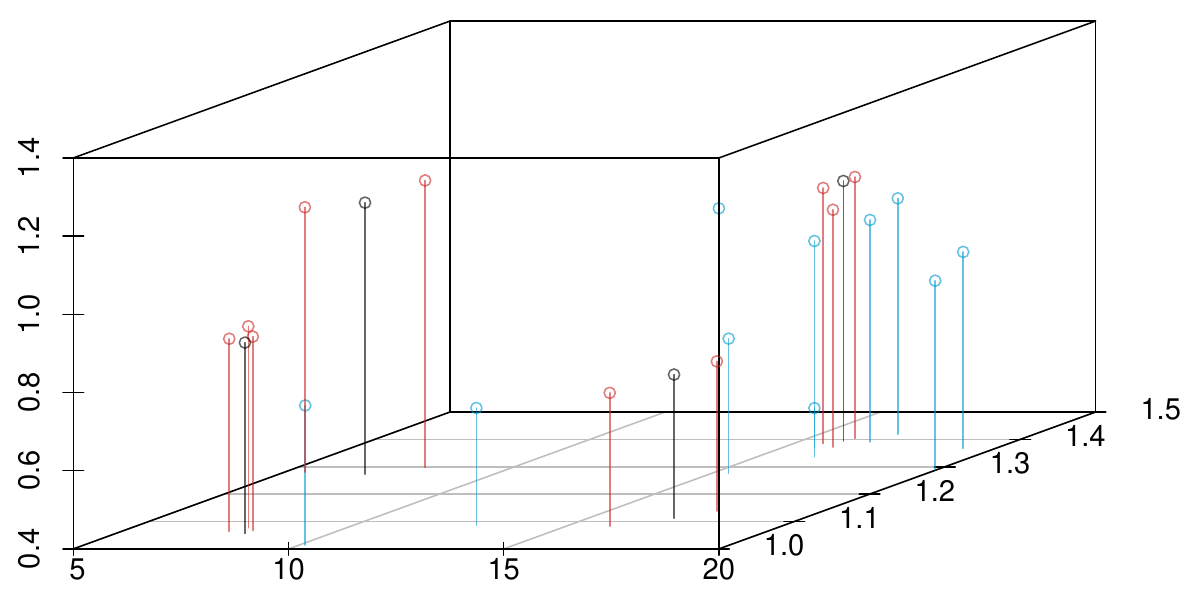}
		\subcaption{$X_{n+1} = 45$}
	\end{subfigure}
	\vspace{0.2cm}
	\caption{Simulation of the parameter $\tau_{i, j, x}$ ($x$-axis), $h_{i, j, x}$ ($y$-axis), and $\rho_{i, j, x}$ ($z$-axis). Black dots represent real data, while blue dots are randomly generated points from the kernel density estimation of $f_{i, j, x}$. Red points in the image (b) indicate the augmented bootstrapped data. For all images, $i = -1$, $j = 0$, and $\ell = 0.02$.}
	\label{fig:parameters}
\end{figure}

The remaining parameter to be estimated is the volatility term $\sigma_{i, j, x}$.
Recall that the BBs in \eqref{eq:BB1} and \eqref{eq:BB2} share the same volatility. The Gaussian and Markovian properties of the BB make it easy to come up with the following formula for the maximum likelihood estimator of $\sigma_{i, j, x}$:
\begin{align*}
    \wh{\sigma}_{i, j, x} &:= \sqrt{\frac{1}{M_{i, j, x}}\sum_{m = 1}^{M_{i, j, x}}(u_m - u_{m-1})^{-1}\lrp{Y_{i, j, x}(u_m)\frac{\tau_{i, j, x}(m) - u_{m-1}}{\tau_{i, j, x}(m)  - u_m} - B_{i, j, x}(u_{m-1})}}, \\
\end{align*}
with $M_{i, j, x} := \sum_{k=1}^{x} \Ind(Y_{i, j, x}(k) = E_{i, j, x}(t_k))$, and
\begin{align}
    u_0 &= 0,\quad u_m = \min\lrc{ k : k > u_{m-1}, Y_{i, j, x}(k) = E_{i, j, x}(k)},\quad m = 1, \dots, M_{i, j, x}, \label{eq:efective_time} \\
    \tau_{i, j, x}(m) &= \Ind(u_m \leq \tau_{i, j, x})\tau_{i, j, x} + \Ind(u_m > \tau_{i, j, x})(x + 1). \label{eq:horizon}
\end{align}
The definition of $u_m$ in \eqref{eq:efective_time} is necessary to pick up only the times that truly represent the jumps in the BBs' paths, and do not account for spurious values due to the representation of the error process $E_{i, j, x}$ in \eqref{eq:error-process-representation}. In \eqref{eq:horizon}, we set $\tau_{i, j, x}(m)$ as a single term to denote the different terminal points of $B_{i, j, x}^{(1)}$ and $B_{i, j, x}^{(2)}$.

One might be tempted to assume that $\sigma_{i, j, x}$ is homogeneous with respect to other parameters like $\rho_{i, j, x}$, $\tau_{i, j, x}$, $h_{i, j, x}$, and $x$, or the battery stages $J_n = i$ and $J_{n+1} = j$. However, empirical evidence suggests otherwise. For instance, higher heights $h_{i, j, x}$ tend to produce higher volatilities, as Figure \ref{fig:vol_vs_height} shows. It also illustrates that a convenient transformation of the volatility might result in a linear relationship between these two parameters.

\begin{figure}[!ht]
	\centering
	\begin{subfigure}[b]{0.49\textwidth}
		\includegraphics[width = \textwidth]{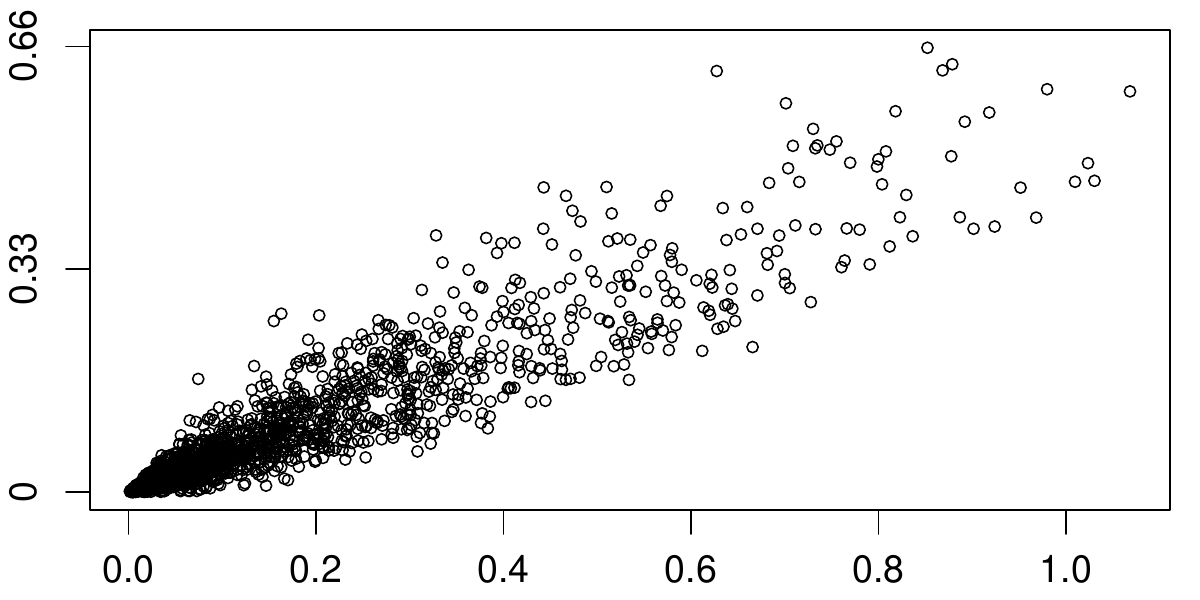}
		\subcaption{$i = -1$ and $j = 0$}
	\end{subfigure}
	\begin{subfigure}[b]{0.49\textwidth}
		\includegraphics[width = \textwidth]{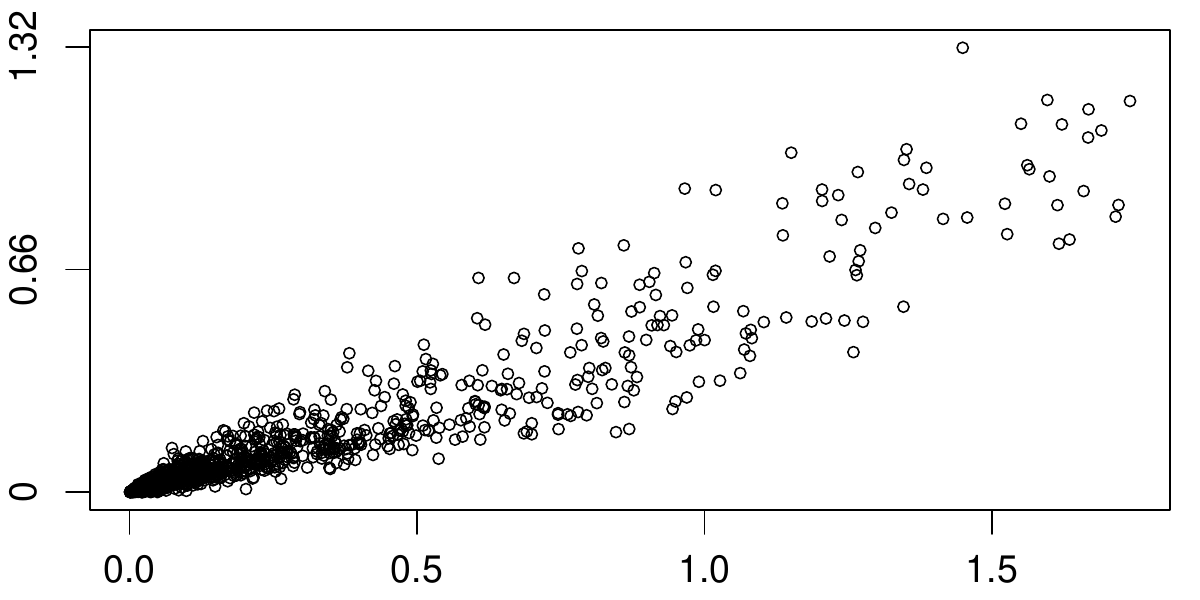}
		\subcaption{$i = +1$ and $j = 0$}
	\end{subfigure}
	\vspace{0.2cm}
	\caption{Images of the relation between the volatility and the height. The $y$ axis marks the values of $\wh{\sigma}_{i, j, x}$ for any value $x\in\N$, and with $i = -1$ and $j = 0$ for figure (a), and $i = -1$ and $j = 0$ for figure (b), while values of $h_{i, j, x}$ are in the $x$-axis. For both figures, $\ell = 0.02$.}
	\label{fig:vol_vs_height}
\end{figure}

In light of this numerical evidence, we take $\rho_{i, j, x}$, $\tau_{i, j, x}$, $h_{i, j, x}$, $x$, and all their first-order interactions, as regressors in a linear model where a transformation of $\wh{\sigma}_{i, j, x}$ is the response. The transformation is chosen from a catalog of several parameterized functions, such that it better helps the linear model to meet its assumptions, namely normality, homoscedasticity, and linearity. We used the \texttt{R} package \texttt{trafo} \citep{trafo_package} to perform this transformation selection. All cases pointed out to Box-Cox-type transformations, having the following form
\begin{align*}
    q_\lambda(x) = 
    \begin{cases}
    (x^\lambda - 1)/\lambda, & \lambda\neq 0 \\
    \log(x), & \lambda = 0
    \end{cases}.
\end{align*}

The final estimation of the volatility $\sigma_{i, j, x}$ is taken to be the anti-transformed mean of the linear model response, whose parameters are chosen to better fit the data
\begin{align*}
    \lrc{(\widehat{\sigma}_{J_m, J_{m+1}, X_{m+1}}, \rho_m, \tau_m, h_m, X_{m+1}) : J_m = i, J_{m+1} = j}_{m\in\N}.
\end{align*}
Actually, we fit the linear model twice. A first fitting was used to ditch out observations with outlier Cook's distances, according to Tukey's method of tagging an outlier as anything farther than 3 times the interquartile range from the median. The second and final fitting was done with the remaining observations. Figure \ref{fig:linear_assumptions} illustrates the final fit for $i = -1$ and $j = 0$. Table \ref{tab:r2} shows the values of the adjusted $R^2$ as a metric of the goodness of the different linear models. 

\begin{figure}[H]
	\centering
	\begin{subfigure}[b]{0.49\textwidth}
		\includegraphics[width = \textwidth]{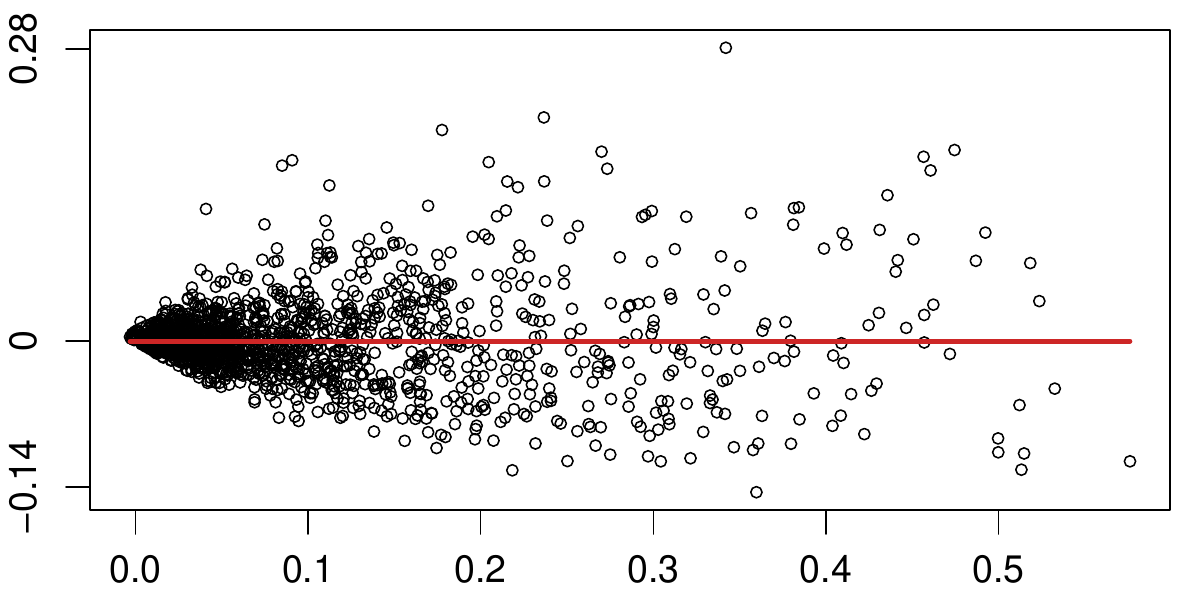}
		\subcaption{Original fitted vs residuals}
	\end{subfigure}
	\begin{subfigure}[b]{0.49\textwidth}
		\includegraphics[width = \textwidth]{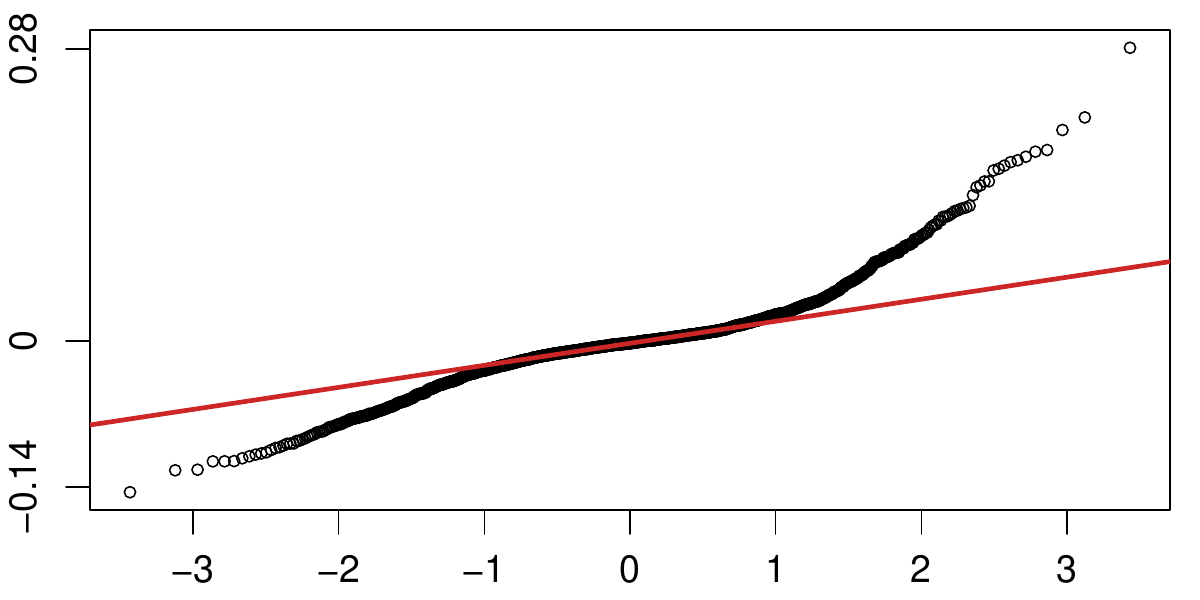}
		\subcaption{Original Q-Q plot}
	\end{subfigure}
	\begin{subfigure}[b]{0.49\textwidth}
		\includegraphics[width = \textwidth]{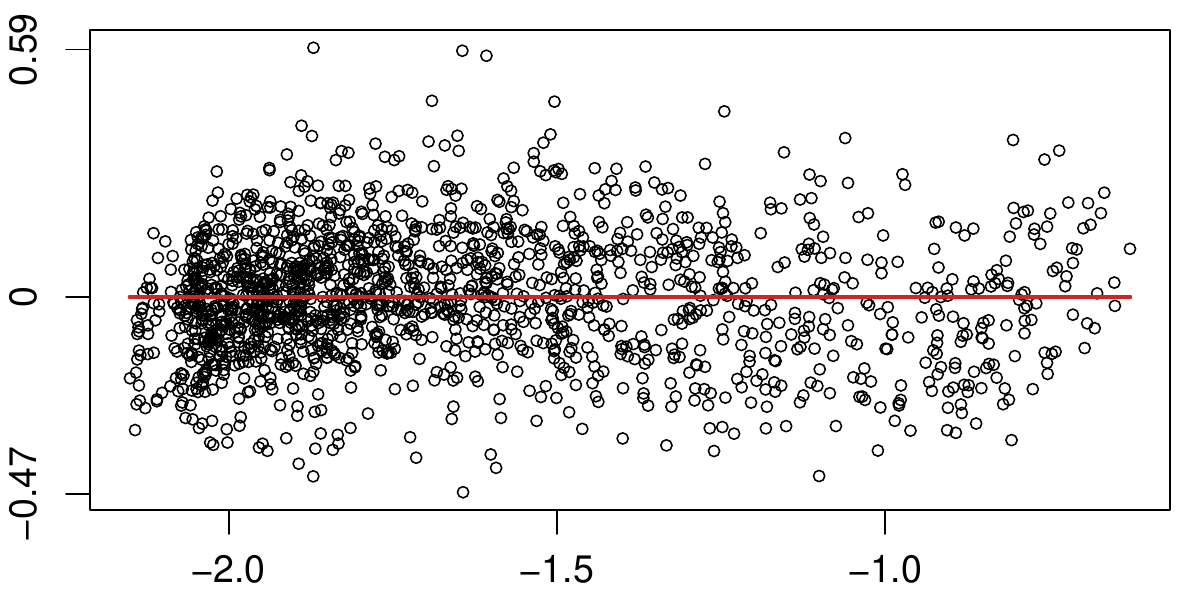}
		\subcaption{Transformed fitted vs residuals}
	\end{subfigure}
	\begin{subfigure}[b]{0.49\textwidth}
		\includegraphics[width = \textwidth]{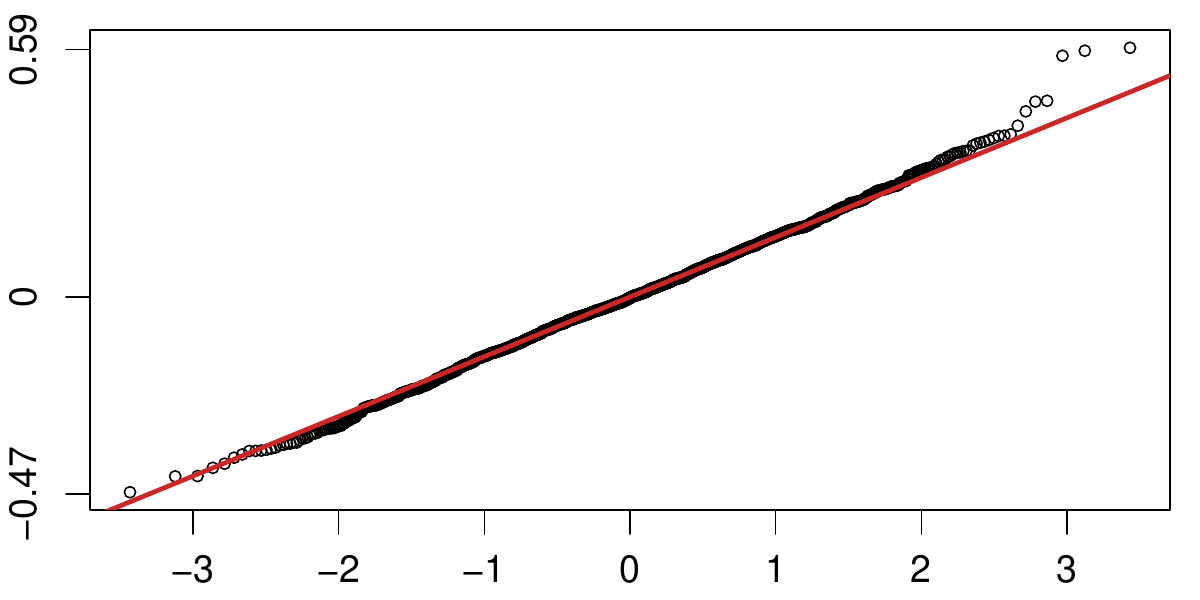}
		\subcaption{Transformed Q-Q plot}
	\end{subfigure}
	\vspace{0.2cm}
	\caption{Check for the linear-model assumptions before and after taking the transformation $q_\lambda$, with $i = -1$, $j = 0$, and $\ell = 0.02$.}
	\label{fig:linear_assumptions}
\end{figure}

\begin{table}[H]
\centering
\begin{tabular}{r|ccc}
                    & $\ell = 0.02$  & $\ell = 0.1$  & $\ell = 0.14$ \\ \hline
$i = -1, j = 0$     & 0.8607      & 0.8127      & 0.8013  \\
$i = -1, j = +1$    & 0.8534      & 0.7591      & 0.7251  \\
$i = +1, j = -1$    & 0.9068      & 0.7415      & 0.7764  \\
$i = +1, j = 0$     & 0.9061      & 0.8295      & 0.7795
\end{tabular}
\caption{Adjusted R$^2$ of the different linear models}
\label{tab:r2}
\end{table}

The GitHub repository \url{https://github.com/aguazz/WindPower-BatteryCharge} provides all the \texttt{R} code and data necessary to implement the estimations and numerical algorithms introduced in this section.

Section \ref{sec:simulations} provides an algorithm to simulate paths of $\cC_{i, j, x}$. The algorithm's performance is validated by comparing the mean and covariance matrices of real and simulated data of $\cC_{i, j, x}$, for different values of $i$, $j$, and $x$.

%-------------------------------------------------%
\section{Techno-economical analysis}\label{sec:SoC_penalty}
%-------------------------------------------------%

The study of the ramp-rate policy requires an analysis of the battery's State Of Charge (SOC) along with the mechanism of the penalty cost. 

Consider the backward-recurrence-time process $B(k) = k - K_{N(k)}$, and let $S(k)$ represents the SOC of the battery at time $k \geq 1$, defined by
\begin{align*}
    S(k) &:= 
    \begin{cases}
        (C_{N(k)}(B(k) + 1) + S(k-1)) \land \ol{c}, & \text{if } J_{N(k)} = +1 \\
        (S(k-1) - C_{N(k)}(B(k) + 1)) \lor \ul{c}, & \text{if } J_{N(k)} = -1 \\ 
        S(k-1), & \text{if } J_{N(k)} = 0
    \end{cases},
\end{align*}
where $\ol{c}$ and $\ul{c}$ are the maximum and minimum SOC levels, respectively. Note that $S(k)\in[\ul{c},\ol{c}]$ for all $k \geq 1$. We remind that, although it is not explicitly stated in the equation above, $C_n$ depends on $J_n$, $J_{n+1}$, and $X_{n+1}$. The previous relation is illustrative of the nonlinear nature of the considered stochastic system. The state of charge process is the result of a nonlinear transformation applied to the charging/discharging process which involves the random charge/discharge and the limit of the battery's capacity.
It is obvious that, in contrast to linear reward structures, the penalty process inhales the nonlinearity of the SOC process and makes it challenging to evaluate the accumulated discounted penalty process.

Likewise, consider the penalty process
\begin{align*}
    M(k) &:= 
    \begin{cases}
        x_{+1}(C_{N(k)}(B(k) + 1) - (\ol{c} - S(k-1))^+, & \text{if } J_{N(k)} = +1 \\
        x_{-1}(C_{N(k)}(B(k) + 1) - (S(k-1) - \ul{c}))^+, & \text{if } J_{N(k)} = -1 \\ 
        0, & \text{if } J_{N(k)} = 0
    \end{cases},
\end{align*}
where the constants $x_{+1}$ and $x_{-1}$ are the penalties per unit of time associated with up-ramping and down-ramping events, respectively.

Finally, consider the cumulative discounted penalty up until time $k \in \N$, defined as
\begin{align*}
W(k) := \sum_{m=0}^k M(m) e^{-r m},
\end{align*}
where $r \geq 0$ is the discount rate.

%-------------------------------------------------%
\section{System simulation}\label{sec:simulations}
%-------------------------------------------------%

The following two simulation algorithms can be used to generate random paths of the charging/discharging process $C_{i, j, x}$ discussed in Section \ref{sec:GM_model}, as well as the SOC and the penalty processes, $S$ and $M$, introduced in Section \ref{sec:SoC_penalty}.
\vspace{0.3cm}

\begin{algorithm}[H]
    \textbf{Input: } $i\in\{-1, 0, +1\}$, $j\in\{-1, 0, +1\}\backslash\{i\}$, $x\in\N$ \\
    \textbf{Output: } $\{c_n\}_{n = 0}^{x+1}$ \\ \vspace{0.1cm}
    \textbf{Pseudo-Code:}\\
        Set $c_{0} = 0$, $c_{x+1} = 0$ \\
        Compute simulations $\tau$, $h$, and $\rho$, from $\tau_{i, j, x}$, $h_{i, j, x}$, $\rho_{i, j, x}$, according to the equations (\ref{eq:max_parameter})~and~(\ref{eq:initial_power}) and the mechanism provided in Section \ref{subsec:parameters} \\
        Simulate two independent standard Brownian motion paths, $\{w_k^{(1)}\}_{k = 0}^{\tau}$ and $\{w_k^{(2)}\}_{k = 0}^{x+1-\tau}$ \\
        Simulate the two independent BB processes $\{y_k^{(1)}\} := \{w_k^{(1)} - (k/\tau)w_{\tau}^{(1)}\}_{k = 0}^{\tau}$ and $\{y_k^{(2)}\} := \{w_k^{(2)} - (k/(x+1-\tau)w_{\tau}^{(2)}\}_{k = 0}^{x+1-\tau}$ \\
	\For{$k = 1$ \KwTo $x$}{
            Set $y_k := y_k^{(1)}\Ind(k \leq \tau) + b_k^{(2)}\Ind(k > \tau)$ as in \eqref{eq:BB} \\
            Compute the triangle process $g_k := g_{i, j, x}(k)$ as in \eqref{eq:triangle} \\
            Compute the error process $e_k := \max\lrc{-g_k, \min\lrc{b_k), \rho - (k - 1)\ell - g_k)}}$ as in \eqref{eq:error-process-representation} \\
            Compute the charge process $c_k := e_k + g_k$ according to \eqref{eq:error-process}
        }
        \caption{Battery charge/discharge simulator}
	\label{alg:charge_simulator}
\end{algorithm}
\vspace{0.3cm}

Figure \ref{fig:real-vs-simulated-charges} shows simulated paths of $\cC_{i, j, x}$, for different values of $i$, $j$, and $x$, produced by implementing Algorithm \ref{alg:charge_simulator}. Note that they visually resemble the real-data paths.

\begin{figure}[!ht]
	\centering
	\begin{subfigure}[b]{0.49\textwidth}
		\includegraphics[width = \textwidth]{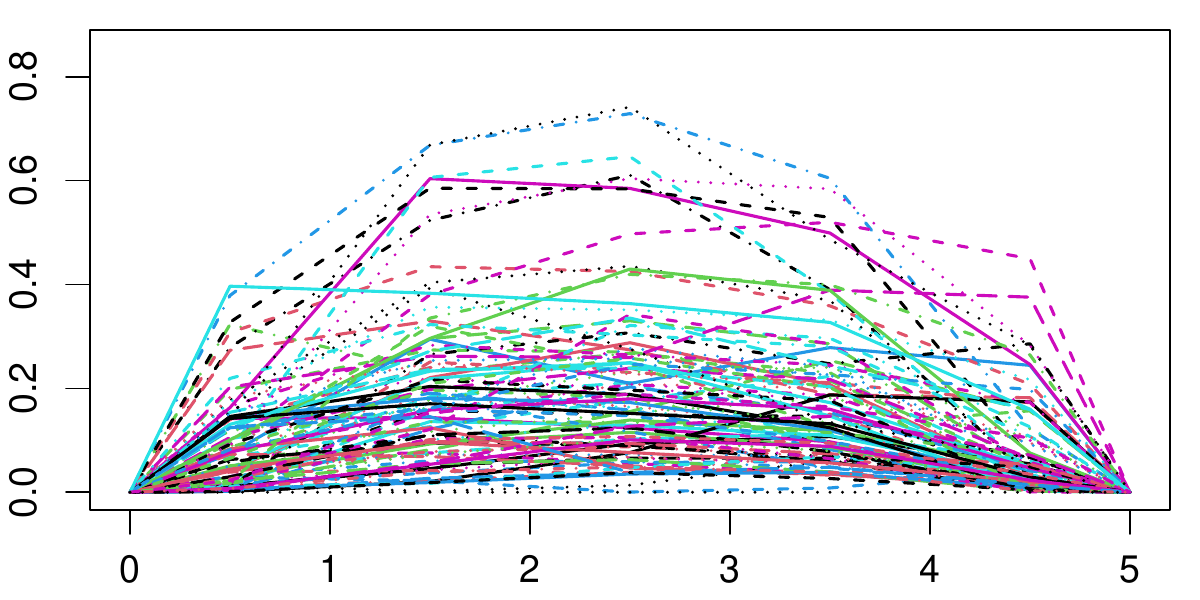}
		\subcaption{}
	\end{subfigure}
	\begin{subfigure}[b]{0.49\textwidth}
		\includegraphics[width = \textwidth]{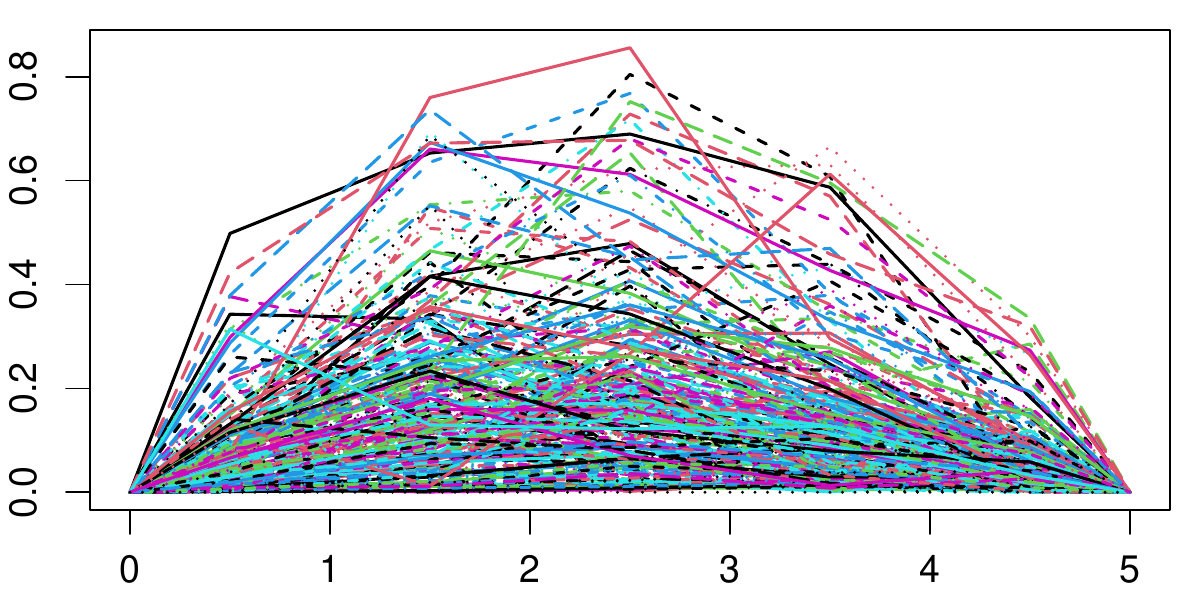}
		\subcaption{}
	\end{subfigure}

        \begin{subfigure}[b]{0.49\textwidth}
		\includegraphics[width = \textwidth]{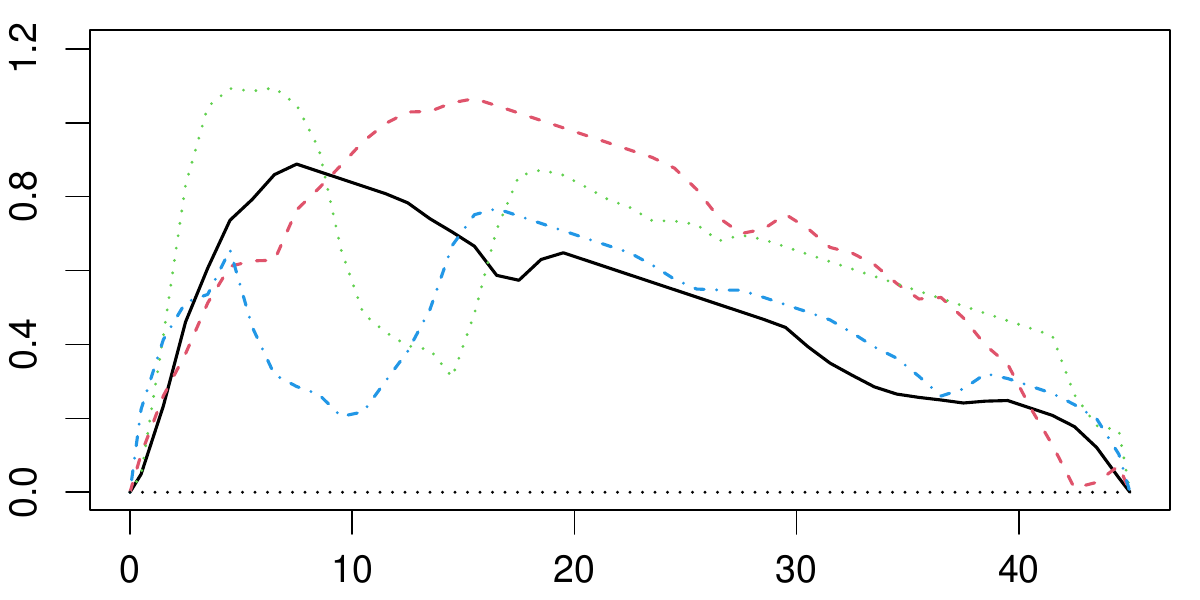}
		\subcaption{}
	\end{subfigure}
	\begin{subfigure}[b]{0.49\textwidth}
		\includegraphics[width = \textwidth]{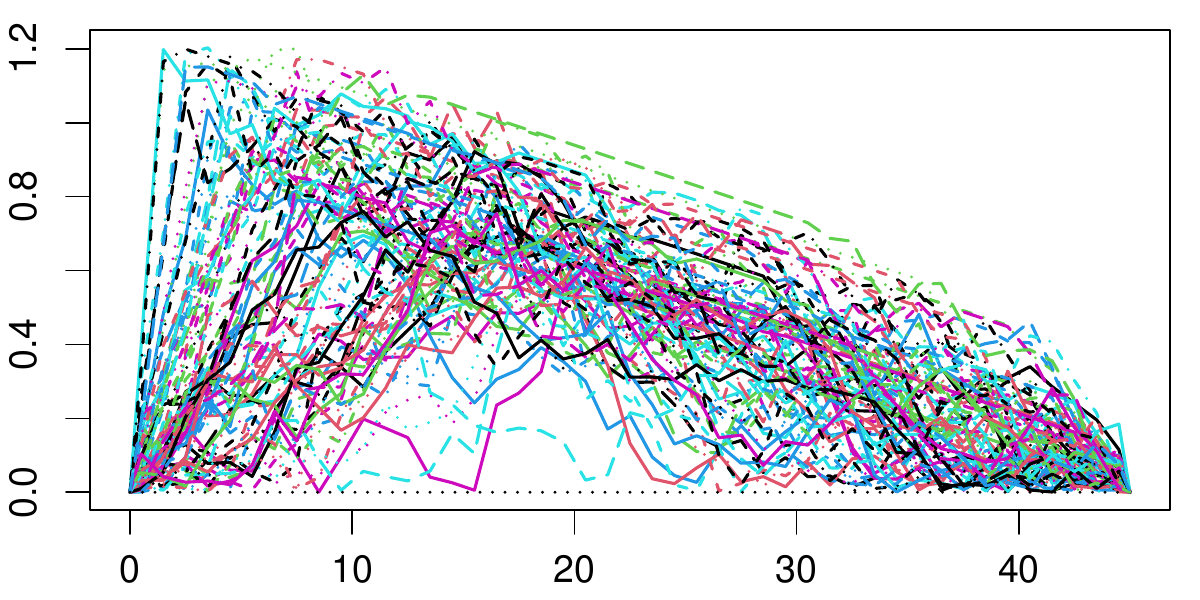}
		\subcaption{}
	\end{subfigure}
	\vspace{0.2cm}
	\caption{Real versus simulated charges. The first column of images accounts for real-data charges, that is, paths of $\cC_{i, j, x}$ pinned to $(0, 0)$ and $(x + 1, 0)$. The second column displays simulated paths of $\cC_{i, j, x}$. The images on the first raw account for $x = 5$, while the second row uses $x = 45$. For all images, $i = -1$, $j = 0$, and $\ell = 0.02$.}
    \label{fig:real-vs-simulated-charges}
\end{figure}

Besides the visual validation in Figure \ref{fig:real-vs-simulated-charges}, we provide the relative mean $L^2$-error between the real and the estimated sample means and sample covariance matrices, displayed in figures \ref{fig:mean-error} and  \ref{fig:cov-error}, respectively. 
We only considered those random segments with at least $30$ observations. For each random segment, we simulated as many paths as the maximum between $3$ times the real-data sample size and $100$ trajectories.

\begin{figure}[H]
	\centering
	\begin{subfigure}[b]{0.49\textwidth}
		\includegraphics[width = \textwidth]{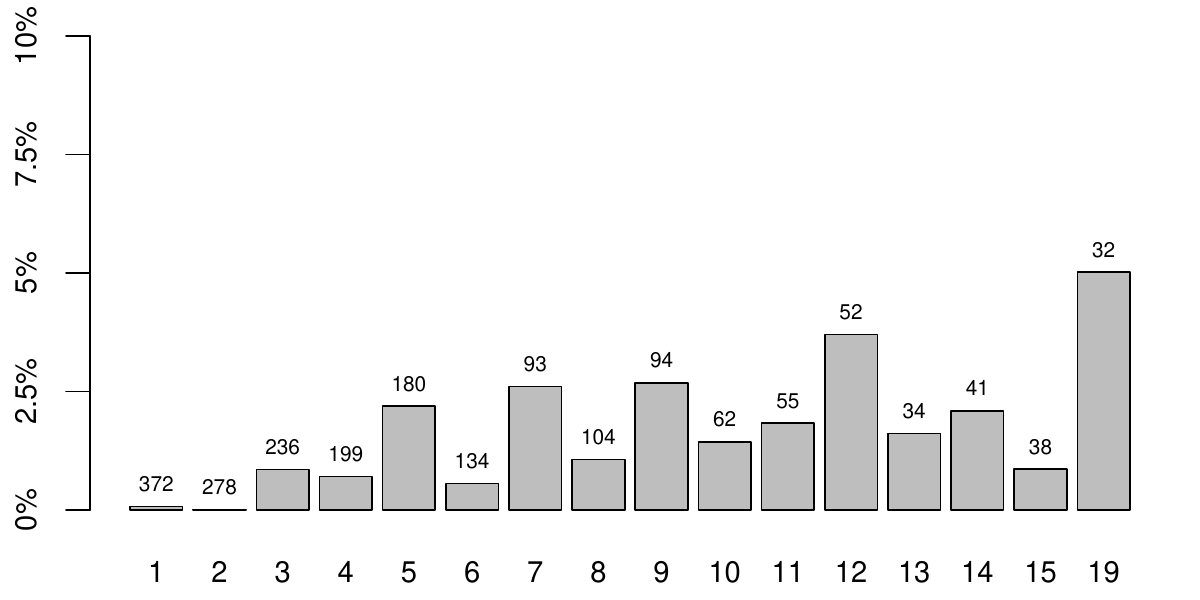}
		\subcaption{$(J_n, J_{n+1}) = (-1, 0)$}
	\end{subfigure}
	\begin{subfigure}[b]{0.49\textwidth}
		\includegraphics[width = \textwidth]{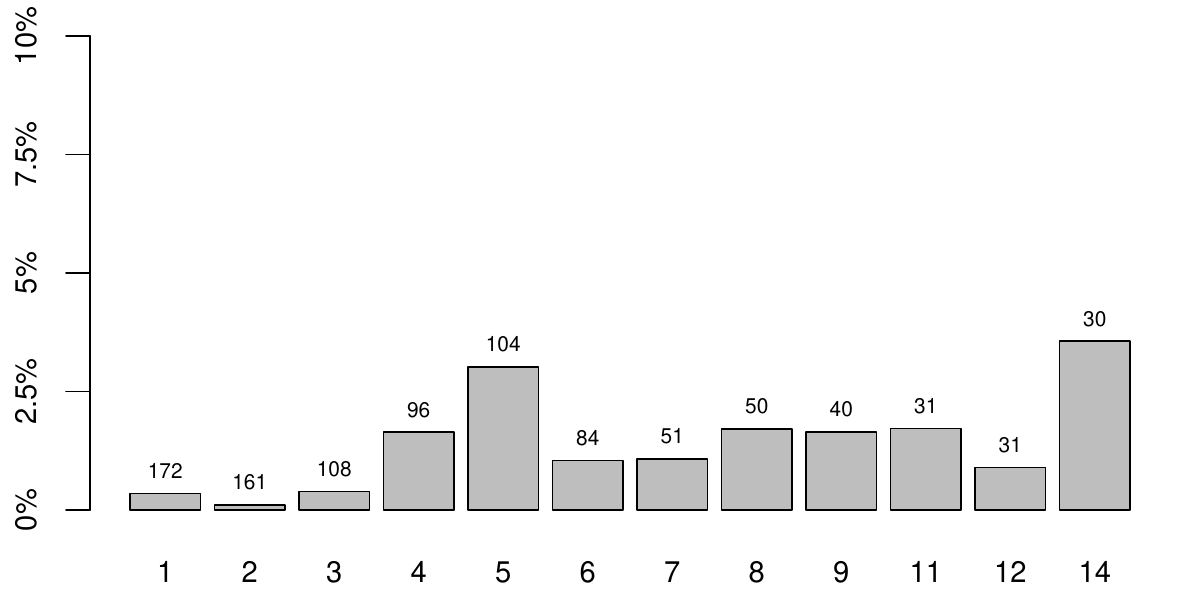}
		\subcaption{$(J_n, J_{n+1}) = (-1, +1)$}
	\end{subfigure}

        \begin{subfigure}[b]{0.49\textwidth}
		\includegraphics[width = \textwidth]{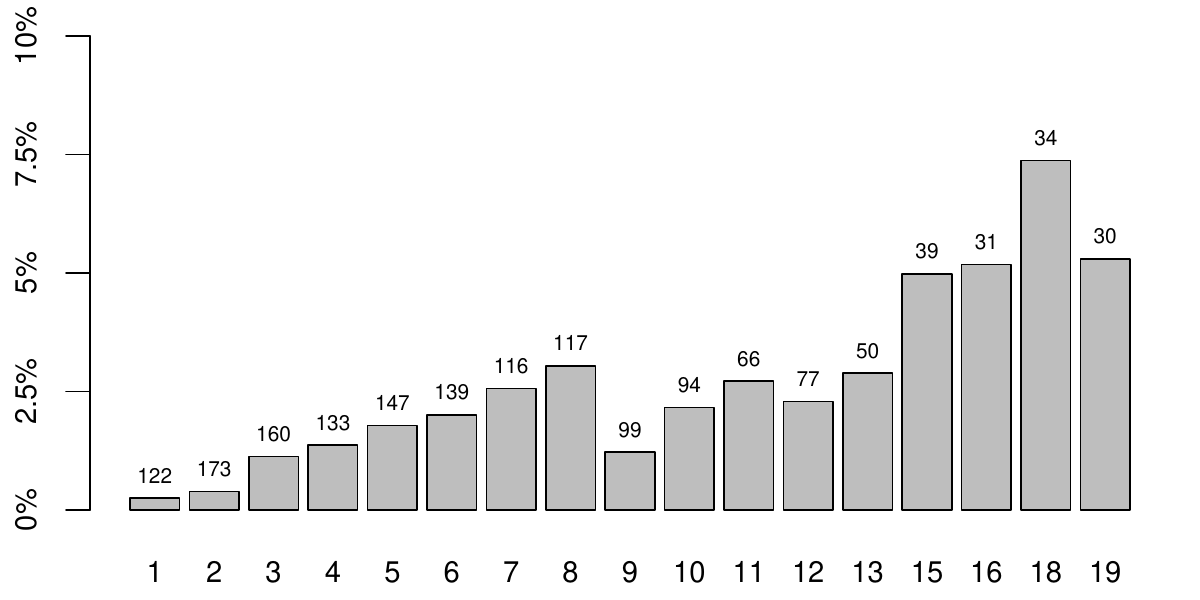}
		\subcaption{$(J_n, J_{n+1}) = (+1, -1)$}
	\end{subfigure}
	\begin{subfigure}[b]{0.49\textwidth}
		\includegraphics[width = \textwidth]{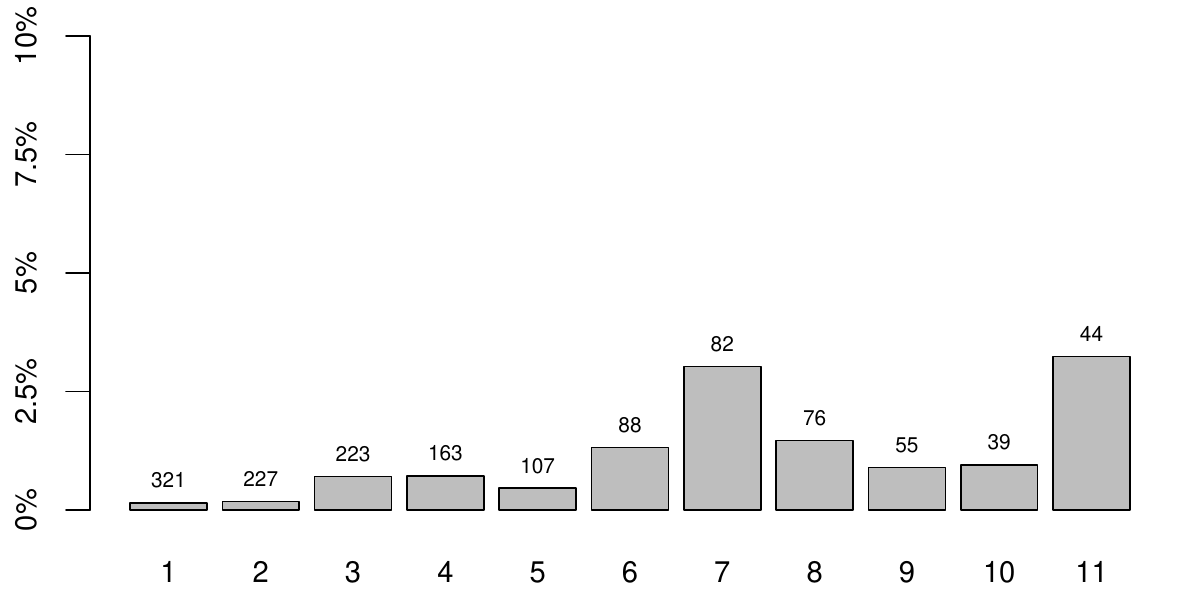}
		\subcaption{$(J_n, J_{n+1}) = (+1, 0)$}
	\end{subfigure}
	\vspace{0.2cm}
	\caption{Relative $L^2$-error, expressed in percentage terms, between the real-data and simulated sample mean. The numbers in the $x$-axis represent the values of $X_{n+1}$, while the numbers on top of the bars are the real-data sample size. For all the images, $\ell = 0.02$.}
    \label{fig:mean-error}
\end{figure}

\begin{figure}[H]
	\centering
	\begin{subfigure}[b]{0.49\textwidth}
		\includegraphics[width = \textwidth]{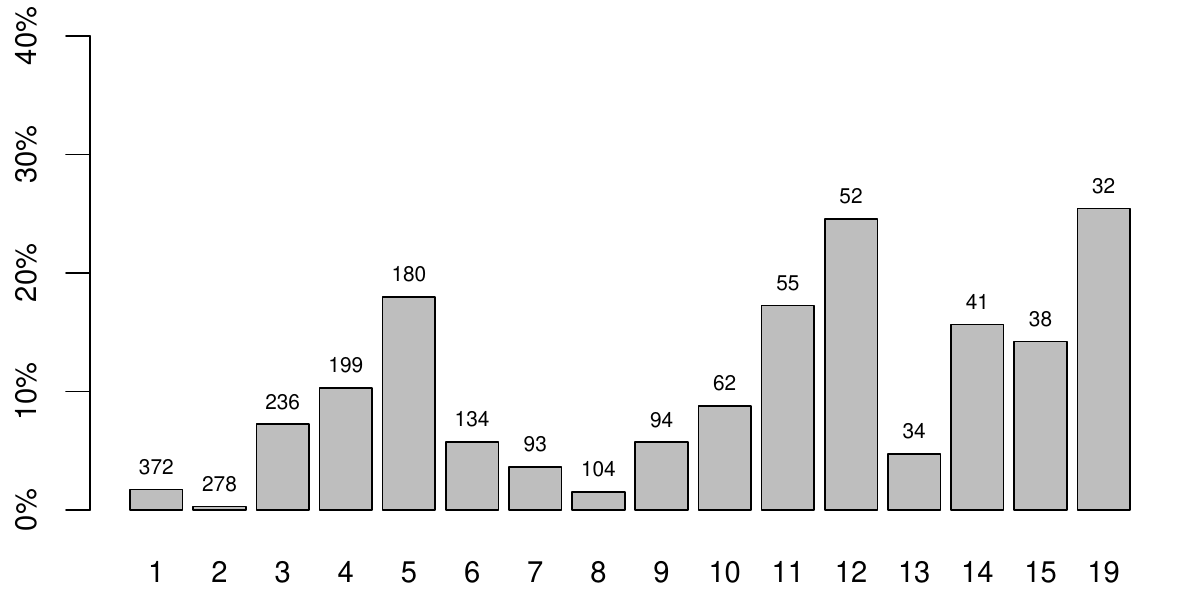}
		\subcaption{$(J_n, J_{n+1}) = (-1, 0)$}
	\end{subfigure}
	\begin{subfigure}[b]{0.49\textwidth}
		\includegraphics[width = \textwidth]{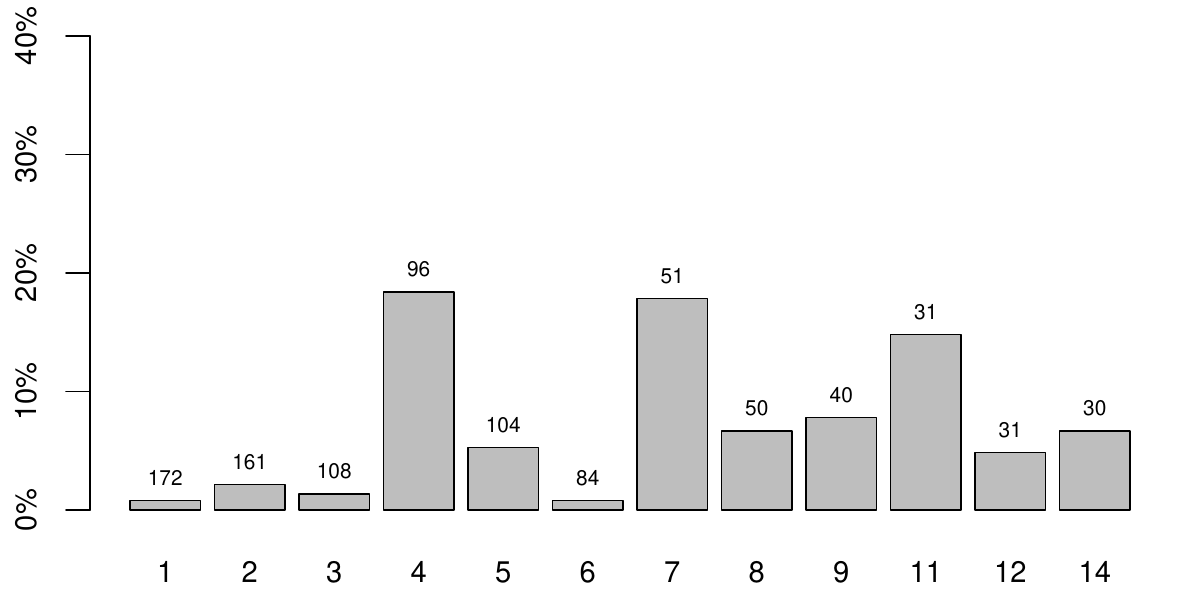}
		\subcaption{$(J_n, J_{n+1}) = (-1, +1)$}
	\end{subfigure}

        \begin{subfigure}[b]{0.49\textwidth}
		\includegraphics[width = \textwidth]{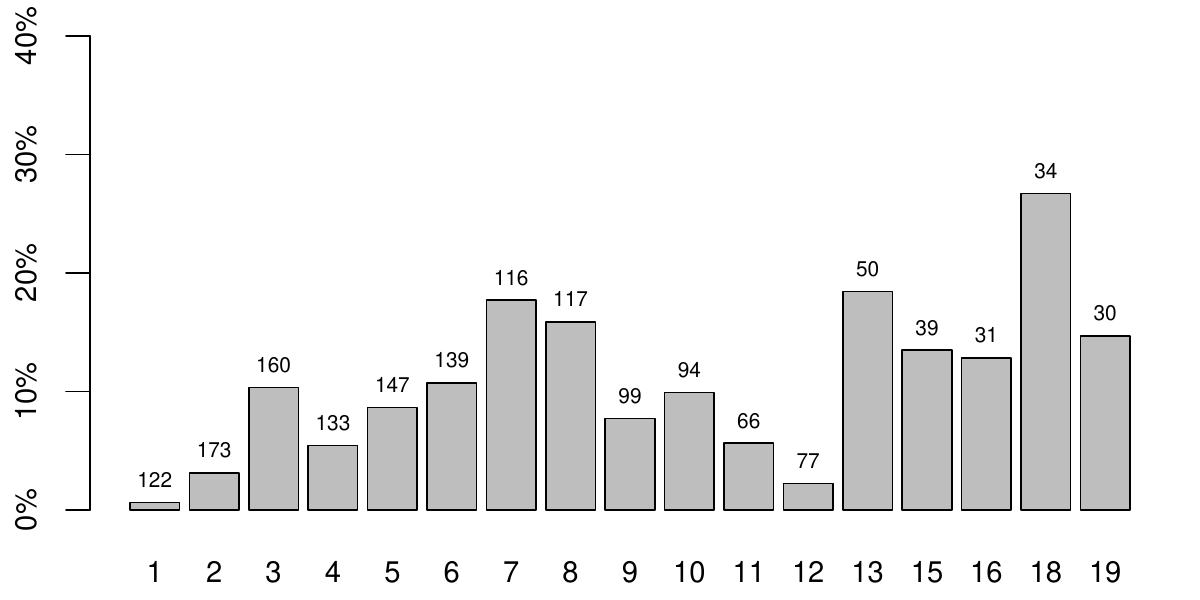}
		\subcaption{$(J_n, J_{n+1}) = (+1, -1)$}
	\end{subfigure}
	\begin{subfigure}[b]{0.49\textwidth}
		\includegraphics[width = \textwidth]{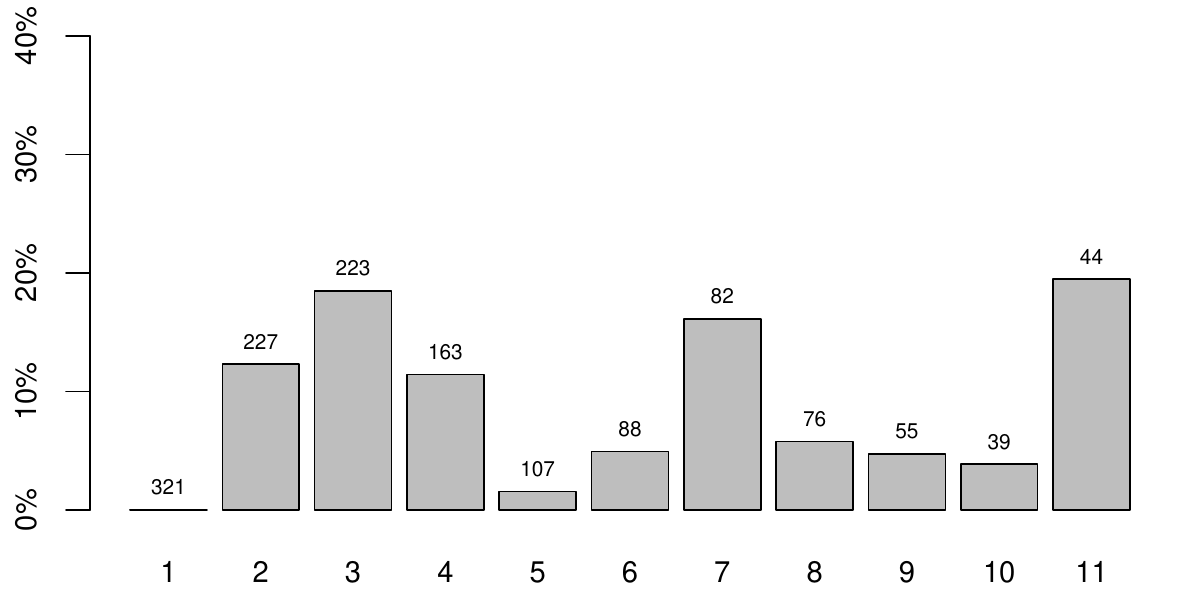}
		\subcaption{$(J_n, J_{n+1}) = (+1, 0)$}
	\end{subfigure}
	\vspace{0.2cm}
	\caption{Relative $L^2$-error, expressed in percentage terms, between the real-data and simulated sample covariance matrices. The numbers in the $x$-axis represent the values of $X_{n+1}$, while the numbers on top of the bars are the real-data sample size. For all the images, $\ell = 0.02$.}
    \label{fig:cov-error}
\end{figure}

Algorithm \ref{alg:penalty_simulator} simulates trayectories of the Markov Renewal chain $\{(J_{n}, K_{n})\}_{n\in\N}$, as well as the SOC process $\{S(k)\}_{k\in\N}$ and the penalty process $\{M(k)\}_{k\in\N}$.

\begin{algorithm}[!ht]
    \textbf{Input: } $N\in\N$, $i\in\{-1, 0, +1\}$, $s\in\R_+$, $b\in\R_+$ \\
    \textbf{Output: } $\{j_n\}_{n = 0}^{N}$, $\{k_n\}_{n = 0}^{N}$, $\{s_k\}_{k = 0}^{k_N}$, $\{m_k\}_{k = 0}^{k_N}$ \\ \vspace{0.1cm}
    \textbf{Pseudo-Code: }\\
        Initialize $j_{0} := i$, $k_0 := 0$, and $s_0 := s$ \\
	\For{$n = 0$ \KwTo $N$}{
            Generate an observation $x$ from the distribution $h_{j_n}$ in \eqref{eq:sojourn-time_density} \\
            Set $k_{n+1} := k_n + x$ \\
            Generate one observation $j$ from the density $p_{ij}(x)$ in \eqref{eq:conditional_transition_prob} \\
            Set $j_{n+1} := j$ \\
            Simulate a path $\lrc{c_k}_{k=0}^{x+1}$ of $\lrc{\cC_{i, j, x}}$, by using Algorithm \ref{alg:charge_simulator} \\
            \lIfElse{$n = 0$}{set $L := b$}{set $L := 0$} 
            \uIf{$i = +1$}{
                \For{$k = 1$ \KwTo $x$}{
                    simulate the SOC process as $s_{k_n + k} = (c_{L + k} + s_{k_n + k - 1}) \land \ol{c}$ \\
                    simulate the penalty process as $m_{k_n + k} = x_{+1}(c_{L + k} - (\ol{c} - s_{k_n + k - 1}))^+$
                }
            }
            \uIf{$i = -1$}{
                \For{$k = 0$ \KwTo $x$}{
                    simulate the SOC process as $s_{k_n + k} = (s_{k_n + k - 1} - c_{L + k}) \lor \ul{c}$ \\
                    simulate the penalty process as $m_{k_n + k} = x_{+1}(c_{L + k} - (s_{k_n + k - 1} - \ul{c}))^+$
                }
            }
            \uIf{$i = 0$}{
                \For{$k = 0$ \KwTo $x$}{
                    simulate the SOC process as $s_{k_n + k} = s_{k_n + k - 1}$ \\
                    simulate the penalty process as $m_{k_n + k} = 0$
                }
            }
	}
	\caption{Semi-Markov reward, SOC, and penalty processes simulator}
	\label{alg:penalty_simulator}
\end{algorithm}
\vspace{0.3cm}

Next we estimate the first and second moments of $W = \{W(k)\}_{k = 1}^K$, for $K = 24$. That is, the hourly average and standard deviation of the cumulative penalty process within a day. To do so, we used Algorithms \ref{alg:charge_simulator} and \ref{alg:penalty_simulator} to simulate $N$ different trayectories of $W$, $W^n = \{W(k)^{(n)}\}_{k = 1}^K$, $n = 1, \dots, N$. Once the N trajectories have been simulated, it is possible to estimate the moments of the accumulated penalty process by computing the corresponding sampling moments

\begin{align*}
\frac{1}{N}\sum_{n=1}^N \lrp{\sum_{k=1}^t M^{(n)}(k)e^{-rk}}^a,\quad a = 1,2,...
\end{align*}

For the wind-farm layout, we consider the battery described in \cite[Table 4]{hittinger2014effect}, that is, a NaS battery with a module energy capacity equal to $0.36$ MWh. These batteries are remarkably cost-efficient compared to super-capacitors and flywheels \citep{hittinger2010compensating}, and their fast response is fundamental to smooth wind-power changes. We consider $10$ years of real-data hourly wind speed to obtain the power production of a hypothetical wind turbine located in Sardinia. As done in \cite{d2022modelling}, we transform the wind speed data into wind power production by means of the function
\begin{align*}
    P(v) := 
    \begin{cases}
        0, & \text{if } v < \ul{v} \\
        P_r\frac{v^3 -\ul{v}^3}{v_r^3 - \ul{v}^3}, & \text{if } \ul{v} < v < v_r \\
        P_r, & \text{if } v_r < v < \ol{v} \\
        0, & \text{if } v > \ol{v}
    \end{cases},
\end{align*}
where $\ul{v}$ is the cut-in wind speed, $\ol{v}$ is the cut-out wind speed, $v_r$ is the rated velocity, and $P_r$ is the rated capacity of the wind turbine. We set $\ul{v} = 4m/s$, $\ol{v} = 25m/s$, $v_r = 13m/s$, and $P_r =2MW$ \citep{d2022modelling,vergine2022optimal}.

The penalty fees are set to $x_{+1} = 21.52$\euro/MWh and $x_{-1} = 26.50$\euro/MWh. These values are the ones used in \cite{hittinger2014effect} after being made proportional to the average electricity price in Italy.

% 1perc_mean 1perc_std
\begin{figure}[!ht]
	\centering
	\begin{subfigure}[b]{0.43\textwidth}
		\includegraphics[width = \textwidth]{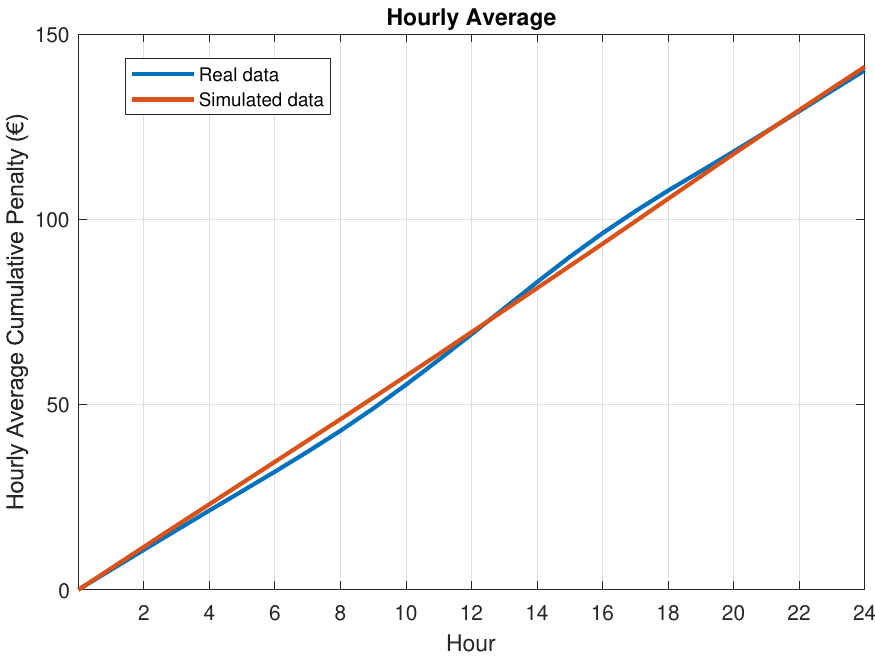}
		\subcaption{Hourly average}
	\end{subfigure}
	\begin{subfigure}[b]{0.43\textwidth}
		\includegraphics[width = \textwidth]{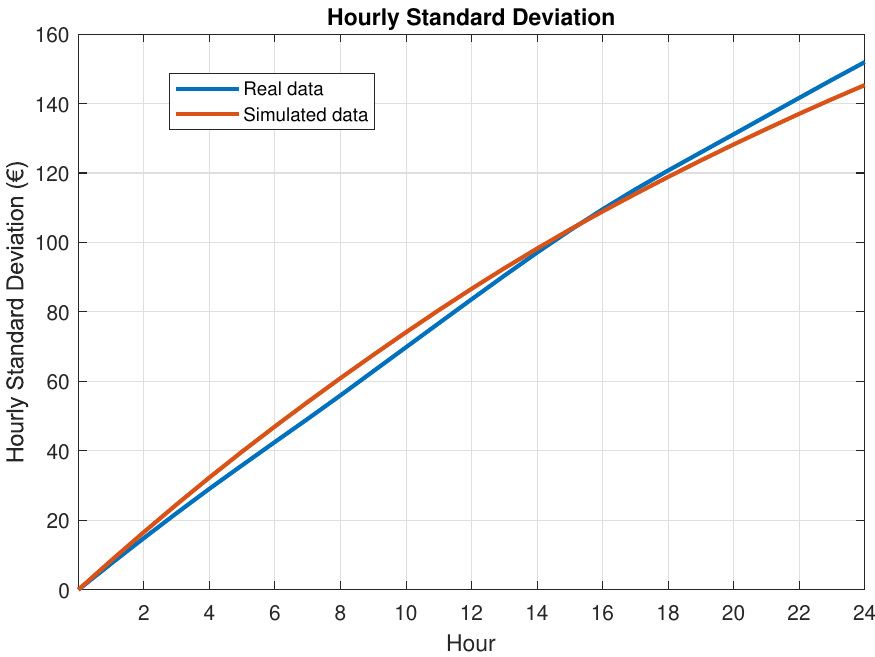}
		\subcaption{Standard deviation}
	\end{subfigure}
	\vspace{6PT}
	\caption{Hourly average cumulative penalty and standard deviation of real and simulated data with ramp-rate limitation percentage equal to $1\%$ ($\ell = 0.02$).}
	\label{fig:ris1}
\end{figure}

% 5perc_mean 5perc_std
\begin{figure}[!ht]
	\centering
	\begin{subfigure}[b]{0.43\textwidth}
		\includegraphics[width = \textwidth]{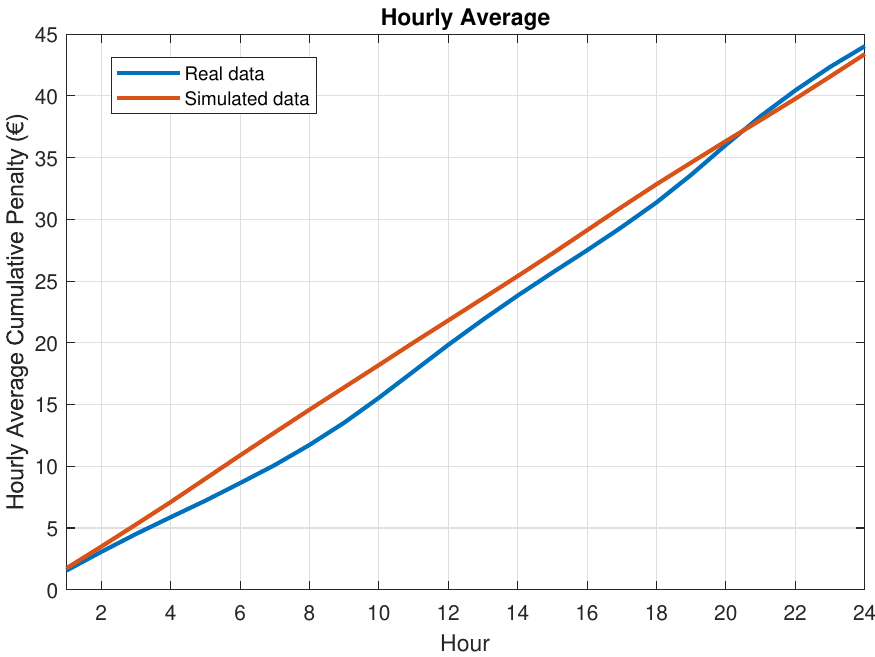} 
		\subcaption{Hourly average}
	\end{subfigure}
	\begin{subfigure}[b]{0.43\textwidth}
		\includegraphics[width = \textwidth]{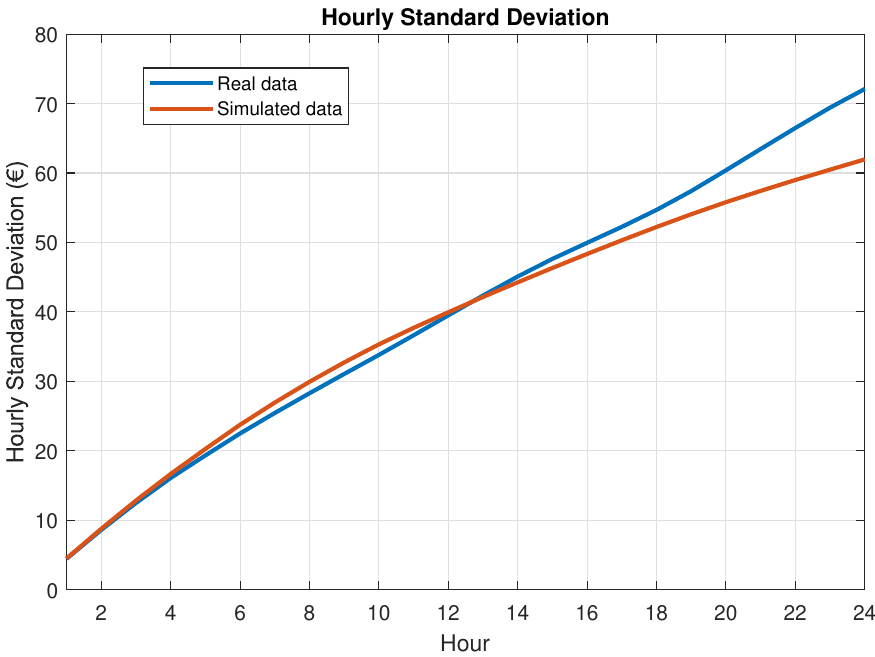} 
		\subcaption{Standard deviation}
	\end{subfigure}
	\vspace{6PT}
	\caption{Hourly average cumulative penalty and standard deviation of real and simulated data with ramp-rate limitation percentage equal to $5\%$ ($\ell = 0.1$).}
	\label{fig:ris2}
\end{figure}

% 7perc_mean 7perc_std
\begin{figure}[!ht]
	\centering
	\begin{subfigure}[b]{0.43\textwidth}
		\includegraphics[width = \textwidth]{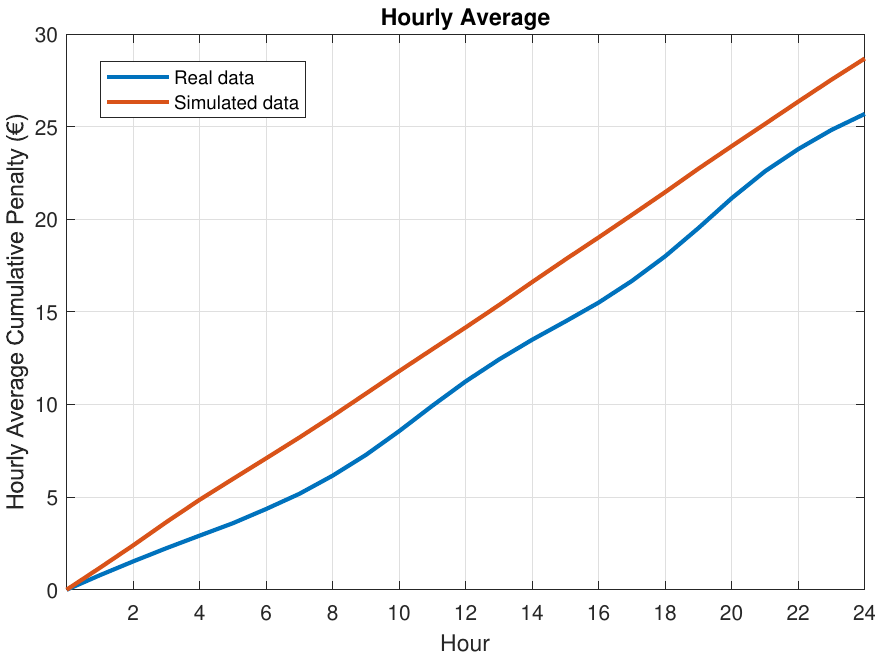} 
		\subcaption{Hourly average}
	\end{subfigure}
	\begin{subfigure}[b]{0.43\textwidth}
		\includegraphics[width = \textwidth]{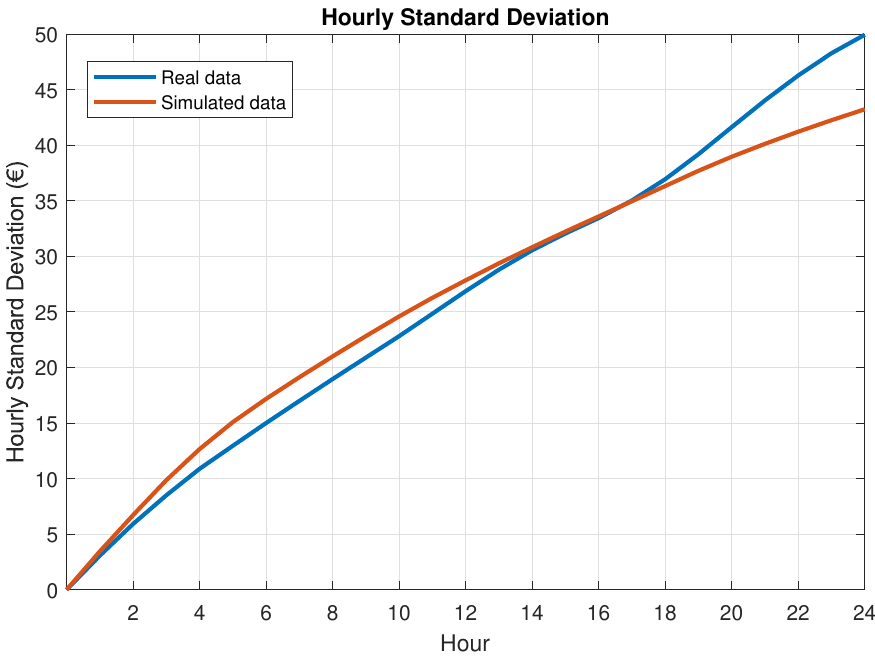} 
		\subcaption{Standard deviation}
	\end{subfigure}
	\vspace{6PT}
	\caption{Hourly average cumulative penalty and standard deviation of real and simulated data with ramp rate limitation percentage equal to $7\%$ ($\ell = 0.14$).}
	\label{fig:ris3}
\end{figure}

The simulations are less accurate for the ramp-rate limitation of $7\%$. This is due to the fact that a higher percentage corresponds to a less strict limitation and, consecutively, to a smaller number of times that the system does not comply with it, which leads to a smaller data-set and, consequently to more biased estimations. 

This fact is supported by the Mean Absolute Percentage Error (MAPE) calculated for the hourly average between real and simulated data. It is a metric that defines how accurate the forecasted quantities are in comparison with the actual quantities and represents the average of the absolute percentage errors. We obtain the values of $2.54$, $11.34$, and $20.18$ for the ramp-rate limitations of $1\%$, $5\%$, and $7\%$, respectively. This confirms the behavior described above with the value of $2.54$, which is very close to the value of $1.77$ obtained in \cite{d2022modelling} with the same ramp-rate limitation, where the proposed model needs a larger number of parameters, being the charge/discharge values independent and not identically distributed at each time withing a sojourn time length. The model proposed in this work gives similar results but captures better the correlation structure in the sample charge/discharge paths. The second-order moment is characterized by greater but contained values of MAPE, $5.56$, $5.92$, and $8.35$ for the three studied limitations.

%-------------------------------------------------%
\section{Concluding remarks}\label{sec:conclusions}
%-------------------------------------------------%

We applied a discrete-time semi-Markov process to model the operations of charge and discharge of a battery storage system connected to a wind farm under a ramp-rate limitation strategy. Within each charging/discharging period, we model the charge/discharge process as a modified Brownian bridge with three parameters.  
The resulting semi-Markov-modulated modified Brownian bridge model was used, via Monte Carlo simulation, to estimate the first and second-order moments of the cumulative discounted penalty coming from violating up-ramp and down-ramp limitations. Not only the estimations are accurate when compared to real data, but they resemble the results obtained by \cite{d2022modelling}, where the authors used a model with a large number of parameters.
In particular, the results show average daily losses ranging from almost $30$\euro\ for a ramp-rate limitation of $7\%$ up to almost $150$\euro\ for the more strict limitation of $1\%$.

Our results can be used to improve the management of the wind farm, since they allow us to obtain detailed information about the state of charge of the battery energy system, as well as the penalty dynamics
associated to a ramp-rate limitation policy. The two algorithms we propose provide an accurate calculation of these variables over time.

Potential extensions of this work include exploring different limitation strategies and storage system technologies and estimating higher moments of the cumulative penalty process. The theoretical calculation of the cumulative penalty process moments is also a worthy path to explore. Finally, using the continuous-time version of the Brownian bridge process in \eqref{eq:BB1} and \eqref{eq:BB2}, one might be able to come up with a continuous-time model for the battery charges/discharges.

This work represents the first step for alluring wind-power producers into accepting ramp-rate policies, by designing effective incentive systems to compensate the potential associated penalties.
These systems have the complementary objective of ensuring the stability of the network by charging costs not only to wind-energy producers.

%-------------------------------------------------%
\section*{Acknowledgments}
%-------------------------------------------------%
Guglielmo D'Amico and Bernardo D’Auria are members of the Gruppo Nazionale Calcolo Scientifico-Istituto Nazionale di Alta Matematica (GNCS-INdAM).

\noindent The authors thank the anonymous referees for their comments, which helped in improving the quality of the manuscript.

%-------------------------------------------------%
\section*{Funding}
%-------------------------------------------------%

This research was partially supported by the Spanish Ministerio de Asuntos Económicos y Transformación Digital grant PID2020-116694GB-I00.

The first author acknowledges the financial support of the Carlos III University of Madrid, by the grant ``Ayuda para la movilidad de investigadores/as de la UC3M en centros de investigación nacionales y extranjeros'', from the program ``Programa Propio de Investigación''.

The last three authors acknowledge the financial support from the program MUR PRIN 2022 n. 2022ETEHRM ‘‘Stochastic models and
techniques for the management of wind farms and power systems’’ by the Italian Ministero dell'Universit\'a e della Ricerca.

The third author also acknowledges the partial financial support from the program MUR PRIN 2022 PNRR n. P20224TM7Z "Probabilistic methods for energy transition" by the Italian Ministero dell'Universit\'a e della Ricerca.

\bibliographystyle{apalike}
\bibliography{BatteryStorage_sM-BB.bib}

\end{document}